\documentclass[sigconf,natbib=true,anonymous=false]{acmart}

\usepackage{booktabs,multirow}
\usepackage{enumitem}
\usepackage{color}
\usepackage{rotating}
\usepackage{amsmath}
\usepackage{array}
\usepackage[linesnumbered,algoruled,boxed,lined]{algorithm2e}

\newcommand{\ie}{{\it i.e.}}
\newcommand{\eg}{{\it e.g.}}
\newcommand{\com}{\textcolor{red}}
\newcommand{\fcom}{\textcolor{blue}}

\newcommand{\ul}{\underline}{}

\newcommand{\maxj}{\max_j}

\usepackage[utf8]{inputenc}

\AtBeginDocument{%
  }

\copyrightyear{2022}
\acmYear{2022}
\setcopyright{acmlicensed}

\acmConference[CIKM '22] {Proceedings of the 31st ACM International Conference on Information and Knowledge Management}{October 17--21, 2022}{Atlanta, GA, USA.}
\acmBooktitle{Proceedings of the 31st ACM International Conference on Information and Knowledge Management (CIKM '22), October 17--21, 2022, Atlanta, GA, USA}
\acmPrice{15.00}
\acmISBN{978-1-4503-9236-5/22/10}
\acmDOI{10.1145/3511808.3557456}
\begin{document}
\title{SpaDE: Improving Sparse Representations using a Dual Document Encoder for First-stage Retrieval}

\author{Eunseong Choi}\authornote{Equal contribution}
\affiliation{
  \institution{Sungkyunkwan University}
  \country{Republic of Korea}}
\email{eunseong@skku.edu}

\author{Sunkyung Lee}\authornotemark[1]
\affiliation{
  \institution{Sungkyunkwan University}
  \country{Republic of Korea}}
\email{sk1027@skku.edu}

\author{Minjin Choi}
\affiliation{
  \institution{Sungkyunkwan University}
  \country{Republic of Korea}}
\email{zxcvxd@skku.edu}

\author{Hyeseon Ko}
\affiliation{
  \institution{Naver Corp.}
  \country{Republic of Korea}}
\email{hyeseon.ko@navercorp.com}

\author{Young-In Song}
\affiliation{
  \institution{Naver Corp.}
  \country{Republic of Korea}}
\email{song.youngin@navercorp.com}

\author{Jongwuk Lee}\authornote{Corresponding author}
\affiliation{
  \institution{Sungkyunkwan University}
  \country{Republic of Korea}}
\email{jongwuklee@skku.edu}

\renewcommand{\shortauthors}{Eunseong Choi et al.}

\date{}
\settopmatter{printacmref=true}

\begin{abstract}
Sparse document representations have been widely used to retrieve relevant documents via \emph{exact lexical matching}. Owing to the pre-computed inverted index, it supports fast ad-hoc search but incurs the \emph{vocabulary mismatch problem}. Although recent neural ranking models using pre-trained language models can address this problem, they usually require expensive query inference costs, implying the trade-off between effectiveness and efficiency. Tackling the trade-off, we propose a novel uni-encoder ranking model, \emph{\textbf{Spa}rse retriever using a \textbf{D}ual document \textbf{E}ncoder (SpaDE)}, learning document representation via the dual encoder. Each encoder plays a central role in (i) adjusting the importance of terms to improve \emph{lexical matching} and (ii) expanding additional terms to support \emph{semantic matching}. Furthermore, our \emph{co-training} strategy trains the dual encoder effectively and avoids unnecessary intervention in training each other. Experimental results on several benchmarks show that SpaDE outperforms existing uni-encoder ranking models.

\end{abstract}

\vspace{-3mm}


\vspace{-2mm}

\maketitle
 
\section{Introduction}
\begin{figure}[t]
\centering
\includegraphics[width=0.9\linewidth]{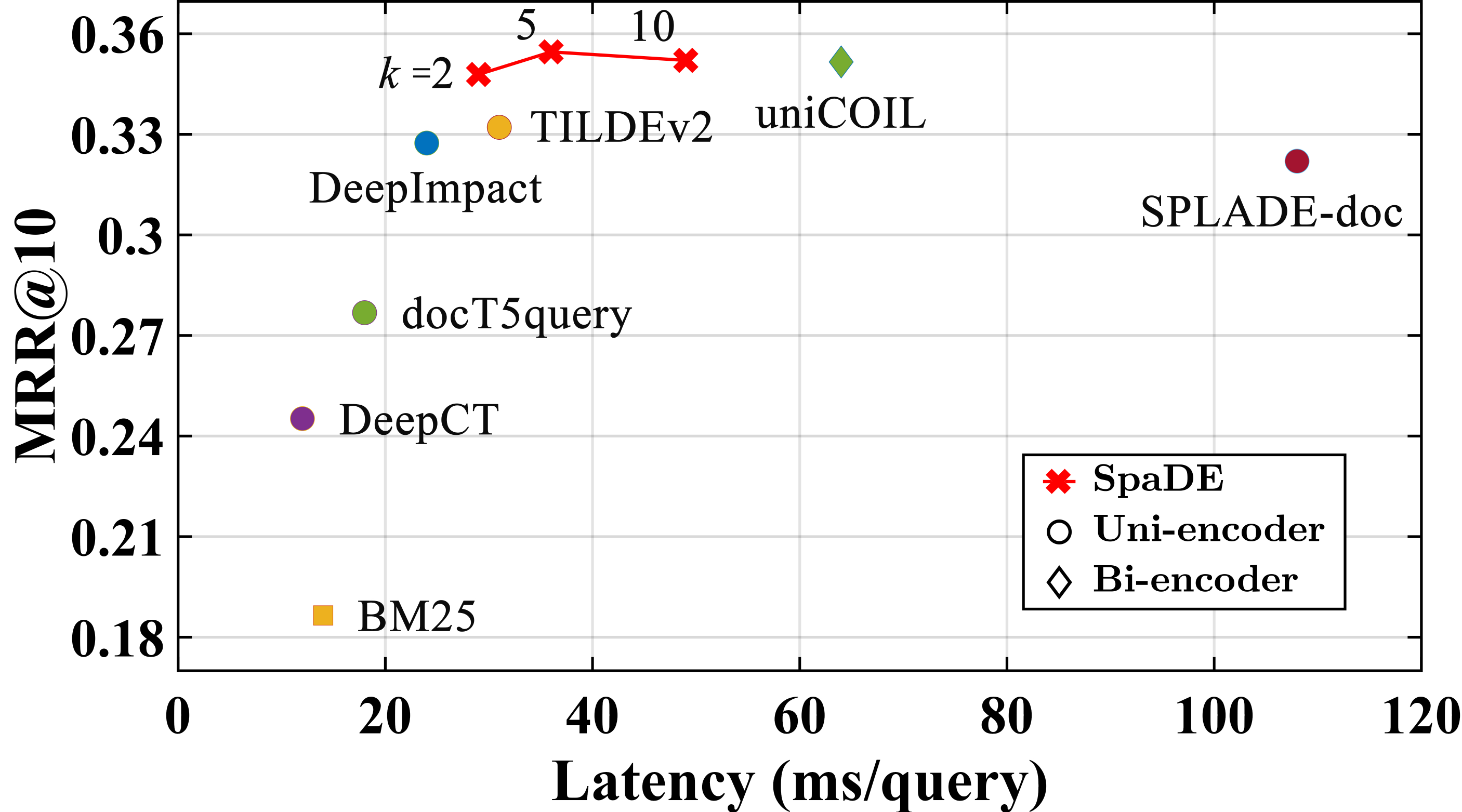}
\vskip -0.1in
\caption{MRR@10 and average query latency (in ms) on MS MARCO development set of SpaDE varying the size of top-$k$ masking for the document-level pruning and existing retrievers with sparse representations. The latency is measured with PISA~\cite{Mallia19PISA} using Block-Max WAND~\cite{DingS11blockmaxwand}.}
\label{fig:expand}
\vskip -0.2in
\end{figure}

Sparse document representations have been commonly used in classical information retrieval (IR) models, such as TF-IDF~\cite{Jones2004TFIDF} and BM25~\cite{RobertsonW94BM25}. Since sparse representations are compatible with the inverted index on the vocabulary space, query inference is remarkably fast for ad-hoc retrieval. However, it merely relies on exact \emph{lexical matching} between a query and a document, inevitably incurring the \emph{vocabulary mismatch problem}, \ie, different terms with similar meanings in queries and documents are mismatched.

Neural ranking models~\cite{ZamaniDCLK18SNRM, GuoFAC16DRMM, XiongDCLP17KNRM, Mitra0C17DUET, HuiYBM17PACRR} have been studied to address the vocabulary mismatch problem. Given mismatched terms, \eg, dog vs. puppy, neural ranking models successfully match their semantics. Notably, pre-trained language models (PLM), \eg, BERT~\cite{DevlinCLT19BERT}, using contextualized representations, have shown remarkable performance leaps on ad-hoc retrieval. As a simple baseline, monoBERT~\cite{NogueiraC19monoBERT} adopts BERT for query-document matching; a \emph{cross-encoder} carries out \emph{all-to-all interactions} across query and document terms, effectively providing complex matching signals. However, it leads to too slow inference time, \eg, > 1,000 ms. Even worse, the cross-encoder model cannot leverage pre-computed indexes, which is impractical for the first-stage retrieval.

Supporting efficient query inference is critical for the first-stage retrieval of billion-scale documents. Recent studies, \eg, ColBERT \cite{KhattabZ20ColBERT, SanthanamKSPZ21ColBERTv2}, COIL~\cite{Gao2021COIL}, and SPLADE~\cite{FormalPC21SPLADE, FormalLPCSPLADEv2}, have proposed to separate query and document encoders, also known as a \emph{bi-encoder}. Given a pair of (query, document), it encodes query and document representations independently and matches them via late interaction. Although the bi-encoder can utilize a pre-computed index for document representations, it is necessary to pass through the query encoder at the retrieval, requiring additional computing costs.

In another research direction, some studies~\cite{DaiC19DeepCT, NogueiraL19docT5query, MalliaKTS21DeepImpact} have designed \emph{uni-encoder} models representing sparse document vectors without a complex query encoder. It is categorized into two directions: (i) \emph{term weighting} to elaborate term scoring or (ii) \emph{term expansion} to mitigate vocabulary mismatching. For term weighting, DeepCT~\cite{DaiC19DeepCT} uses BERT~\cite{DevlinCLT19BERT} to adjust the importance of terms to improve the quality of conventional term frequency scores. However, it still suffers from the vocabulary mismatch problem. For term expansion, doc2query~\cite{NogueiraYLC19doc2query} and docT5query~\cite{NogueiraL19docT5query} generate relevant queries from a document using Transformer~\cite{VaswaniSPUJGKP17Transformer} or T5~\cite{Raffel2020t5} and then add generated queries for document embeddings to enrich document representation. Recently, DeepImpact~\cite{MalliaKTS21DeepImpact} and TILDEv2~\cite{ZhuangZ2021TILDEv2} adjust term scores on the document expanded by docT5query~\cite{NogueiraL19docT5query} or TILDE~\cite{ZhuangZ2021TILDE} respectively, mitigating the limitation of term weighting. While uni-encoder models achieve fast query inference by building the inverted index, they show lower accuracy than cross-encoder and bi-encoder models.

In this paper, we propose a novel uni-encoder ranking model, namely \emph{\textbf{Spa}rse retriever using a \textbf{D}ual document \textbf{E}ncoder (SpaDE)}, to overcome the trade-off between effectiveness and efficiency. Specifically, it adopts a \emph{dual} document encoder advocating two solutions, \ie, \emph{term weighting} and \emph{term expansion}. While term weighting adjusts the term importance scores for lexical matching, term expansion enables semantic matching by appending additional terms to the document. It should be noted that each solution handles a different target scenario, but both scenarios exist in real-world datasets. Although each solution has been used in the literature, it is non-trivial to train the dual encoder effectively.

To address this problem, we devise a simple-yet-effective \emph{co-training} strategy to enjoy the benefit of the dual encoder. Our key idea is to avoid unnecessary intervention in training each encoder with different aspects. Unlike joint training, where the dual encoder is trained with the same samples, our training strategy focuses on selectively training hard samples from another. We choose large-loss samples from one encoder and train another encoder using them, and vice versa. As a result, SpaDE can effectively train the dual encoder with the complementary relationship.

For efficiency, we further leverage a learning-based pruning method to enforce the sparsity of document representations. Using top-\emph{k} masking for \emph{document-level} sparsity and cutoff with approximate document frequency for \emph{corpus-level} sparsity, we gradually prune unnecessary terms in the document and fully enjoy the advantage of the inverted index. Owing to the pruning method, SpaDE exhibits a better trade-off improving query latency by 3.4 times with little accuracy degradation over bi-encoder models.

Figure~\ref{fig:expand} illustrates how well SpaDE alleviates the trade-off between effectiveness and efficiency for the first-stage retrieval on the MS MARCO dataset. It can serve in 29--49 milliseconds when the size of top-$k$ masking is 2--10. Furthermore, it outperforms existing uni-encoder models and significantly reduces the latency with comparable effectiveness to bi-encoders. In this sense, SpaDE can be used as an alternative for the first-stage ranker in commercial search engines.

To summarize, the key contributions of this paper are as follows. (i) We propose a novel ranking model, called SpaDE, for effective and efficient first-stage retrieval (Section~\ref{sec:architecture}). (ii) We introduce a \emph{dual} document encoder, enjoying both term weighting and term expansion for document representations (Section~\ref{sec:dualencoder}). (iii) We also present a \emph{co-training} strategy to effectively train the dual encoder (Section~\ref{sec:modeltraining}) and a learning-based pruning method to build an efficient inverted index (Section~\ref{sec:indexing}). (iv) Lastly, we evaluate the effectiveness and efficiency of the proposed method on various benchmark datasets, such as MS MARCO Passage Ranking~\cite{Nguyen2016msmarco}, TREC DL 2019~\cite{Craswell2020trec2019}, TREC DL 2020~\cite{Craswell2021trec2020}, and BEIR~\cite{ThakurRRSG21BEIR} (Sections ~\ref{sec:setup} and \ref{sec:results}).

\section{Related Work}

\begin{table}[]
\centering
\caption{Category of existing neural ranking models using PLM with two criteria; representation types and matching paradigms. Note that cross-encoder models using sparse representation and uni-encoder models using dense representation do not exist.}\label{tab:existing}
\vskip -0.1in
\begin{footnotesize}
\begin{tabular}{c|c|c}
\toprule
              & \textbf{{\small Sparse representations}}                                                                                                  & \textbf{{\small Dense representations}}                                                            \\ \midrule
\begin{tabular}[c]{@{}c@{}}\textbf{{\small Cross-}}\\\textbf{{\small encoder}}\end{tabular} & -                                                                                                                        & \begin{tabular}[c]{@{}c@{}}monoBERT~\cite{NogueiraC19monoBERT}, \\ duoBERT~\cite{NogueiraCYCK19duoBERT}, \\ Simplified TinyBERT~\cite{ChenHHSS21TinyBERT},\\Birch~\cite{YilmazWYZL19Birch}, CEDR~\cite{MacAvaney2019CEDR}\\BERT-MaxP~\cite{Dai2019bertmaxp},\\PreTTR~\cite{MacAvaney2020PreTTR},\\PARADE~\cite{Li20PARADE}\end{tabular}                 \\ \midrule
\begin{tabular}[c]{@{}c@{}}\textbf{{\small Bi-}}\\\textbf{{\small encoder}}\end{tabular}    & \begin{tabular}[c]{@{}c@{}}EPIC~\cite{MacAvaneyN0TGF20bEPIC}, SparTerm~\cite{BaiLWZSXWWL20SparTerm},\\
uniCOIL~\cite{LinM21uniCOIL}, COIL-tok~\cite{Gao2021COIL}, \\ 
SPLADE~\cite{FormalPC21SPLADE}, SPLADE-max~\cite{FormalLPCSPLADEv2}, \\
DistilSPLADE-max~\cite{FormalLPCSPLADEv2}, \\
UHD-BERT~\cite{JangKHMPYS21UHDBERT}\end{tabular} & \begin{tabular}[c]{@{}c@{}}Sentence-BERT~\cite{ReimersG19SBERT},\\DPR~\cite{KarpukhinOMLWEC20DPR}, ANCE~\cite{XiongXLTLBAO21ANCE},
\\ADORE~\cite{Zhan2021ADORE}, TAS-B~\cite{Hofstatter2021TAS-B},\\Condenser~\cite{Gao2021Condenser, Gao2021coCondenser},\\ ColBERT(v1, v2)~\cite{KhattabZ20ColBERT, SanthanamKSPZ21ColBERTv2},\\TCT-ColBERT~\cite{LinYL20TCT-ColBERT}, JPQ~\cite{ZhanM0GZM21JPQ},\\COIL-full~\cite{Gao2021COIL}, CLEAR~\cite{GaoDFC20CLEAR},\\ CoRT~\cite{WrzalikK21CoRT}, RepCONC~\cite{Zhan22RepCONC},\\RocketQA(v1, v2)~\cite{QuDLLRZDWW21RocketQA, Ren2021RocketQAv2}\end{tabular} \\ \midrule
\begin{tabular}[c]{@{}c@{}}\textbf{{\small Uni-}}\\\textbf{{\small encoder}}\end{tabular}   & \begin{tabular}[c]{@{}c@{}}doc2query~\cite{NogueiraYLC19doc2query},  docT5query~\cite{NogueiraL19docT5query},\\DeepCT~\cite{DaiC19DeepCT}, DeepImpact~\cite{MalliaKTS21DeepImpact},\\SPLADE-doc~\cite{FormalLPCSPLADEv2}\\TILDE(v1, v2)~\cite{ZhuangZ2021TILDE, ZhuangZ2021TILDEv2}\end{tabular}                                      & -                                                                                 \\ \bottomrule
\end{tabular}
\end{footnotesize}
\vspace{-4.5mm}
\end{table}

\begin{figure*}
\centering
\includegraphics[width=1.0\linewidth]{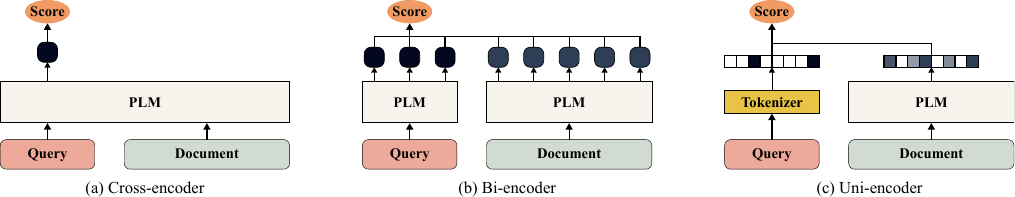}
\vskip -0.05in
\caption{Illustration of neural ranking models using PLM. Existing PLM-based ranking models are categorized into three approaches depending on query-document matching paradigms.}\label{fig:deeplm}
\vskip -0.1in
\end{figure*}

We review existing neural ranking models~\cite{Lin2020Survey} using the pre-trained language model (PLM). As shown in Table~\ref{tab:existing}, they are categorized into \emph{cross-encoder}, \emph{bi-encoder}, and \emph{uni-encoder} (Figure~\ref{fig:deeplm}).

\subsection{Cross-encoder Approach}\label{sec:crossencoder}

This approach regards a query and a document as a pair of sentences and performs all-to-all interactions across query and document terms. First, monoBERT~\cite{ NogueiraC19monoBERT} took the concatenation of the query and the document as input and computed a relevance score as the point-wise ranking problem. Meanwhile, duoBERT~\cite{NogueiraCYCK19duoBERT} formulated the pair-wise ranking problem to classify relevant and irrelevant documents for the query. The two models can be pipelined for a multi-stage ranking in an end-to-end fashion. Birch~\cite{YilmazWYZL19Birch}, BERT-MaxP~\cite{Dai2019bertmaxp}, CEDR~\cite{MacAvaney2019CEDR}, and PARADE~\cite{Li20PARADE} extended passage ranking into document ranking by utilizing BERT’s representations obtained from the cross-encoder. Despite its effectiveness in capturing complex query-document interactions, it is known that query latency is 100-1,000x slower than BM25, as reported in~\cite{KhattabZ20ColBERT}. Recently, to mitigate the computational cost of cross-encoder, PreTTR~\cite{MacAvaney2020PreTTR} stored term representations from an intermediate layer to delay all-to-all interactions, and Simplified TinyBERT~\cite{ChenHHSS21TinyBERT} utilized knowledge distillation models to reduce query inference time.

\vspace{-4.0mm}
\subsection{Bi-encoder Approach}\label{sec:biencoder}
This approach encodes a query and a document separately and learns the interaction between query and document vectors. 

\vspace{1.0mm}
\noindent
\textbf{Dense representations.} The encoder converts the query and the document to continuous low-dimensional vectors, which naturally bypasses the vocabulary mismatch problem. Although Sentence-BERT~\cite{ReimersG19SBERT} does not directly tackle the retrieval problem, it is used as the basic design for the bi-encoder with dense representations. As the canonical example, DPR~\cite{KarpukhinOMLWEC20DPR}, ANCE~\cite{XiongXLTLBAO21ANCE}, and RocketQA~\cite{QuDLLRZDWW21RocketQA, Ren2021RocketQAv2} exploited the [\texttt{CLS}] vectors to compute the relevance score. These models increase effectiveness through more sophisticated negative sampling methods. Recently, ADORE~\cite{Zhan2021ADORE} used dynamic sampling to adapt hard negative samples during model training. TAS-B~\cite{Hofstatter2021TAS-B} proposed topic-aware sampling to select informative queries from topic clusters. Instead of using single dense vector representation, ColBERT~\cite{KhattabZ20ColBERT, SanthanamKSPZ21ColBERTv2} and COIL~\cite{Gao2021COIL} leveraged late interactions between query token vectors and document token vectors.

Meanwhile, Condenser~\cite{Gao2021Condenser, Gao2021coCondenser} designed a pre-training strategy tailored for ad-hoc retrieval, and  RocketQAv2~\cite{Ren2021RocketQAv2}, TCT-ColBERT\\~\cite{LinYL20TCT-ColBERT}, TAS-B~\cite{Hofstatter2021TAS-B}, and ColBERTv2~\cite{SanthanamKSPZ21ColBERTv2} adopted knowledge distillation to improve ranking performance further. To reduce query inference time, dense representation models utilized an approximate similarity search, \eg, \texttt{faiss}~\cite{MalkovY20HNSW, JohnsonDJ21}. However, since the index size for dense vectors is much larger than the traditional inverted index, loading whole dense vectors requires huge memory space. To deal with the memory footprint problem, JPQ~\cite{ZhanM0GZM21JPQ} and RepCONC~\cite{Zhan22RepCONC} compressed dense representations.

\vspace{1.0mm}
\noindent
\textbf{Sparse representations.} Recent studies have represented sparse query and document vectors. EPIC~\cite{MacAvaneyN0TGF20bEPIC} matched a dense document vector and a sparse query vector to reduce the query latency. SparTerm~\cite{BaiLWZSXWWL20SparTerm} adopted sparse document representations by combining the importance distribution of terms with a binary gating vector. Later, SPLADE~\cite{FormalPC21SPLADE} improved SparTerm~\cite{BaiLWZSXWWL20SparTerm} using a FLOPS regularizer~\cite{PariaYYXRP20FLOPS}, enforcing the sparsity of term distributions. SPLADE-max~\cite{FormalLPCSPLADEv2} further improved SPLADE~\cite{FormalPC21SPLADE} by replacing the pooling operation. Although COIL-tok~\cite{Gao2021COIL} and uniCOIL~\cite{LinM21uniCOIL} stem from COIL~\cite{Gao2021COIL}, they adopt sparse representations by excluding \texttt{[CLS]} matching or setting the dimension size of dense vectors as one, respectively. In another direction, UHD-BERT~\cite{JangKHMPYS21UHDBERT} adopted a winner-take-all module for binarized sparse representations in ultra-high dimensional space. Although the bi-encoder design using sparse representations can employ the inverted index, it requires additional costs to compute the query encoder. Therefore, they still have a bottleneck in reducing query inference time.

\vspace{-3.5mm}
\subsection{Uni-encoder Approach}\label{sec:uniencoder}

For efficient query inference, it is preferred to avoid a complex query encoder, as pointed out in \cite{ZhuangZ2021TILDE, ZhuangZ2021TILDEv2}. The uni-encoder models minimize the burden of query inference time, especially enabling practical adoption of GPU-free models. Furthermore, the uni-encoder model is inherently suitable for the inverted index by using only a tokenizer to represent queries into a bag of words.

\vspace{1.0mm}
\noindent
\textbf{Term expansion.} It is reformulated by the machine translation problem. Given a dataset of (query, relevant document) pairs, a sequence-to-sequence model is trained to generate the query from the relevant document. doc2query~\cite{NogueiraYLC19doc2query} and docT5query~\cite{NogueiraL19docT5query} used Transformer~\cite{VaswaniSPUJGKP17Transformer} and T5~\cite{Raffel2020t5} to predict queries and append them to the original document. Expanded documents can also be used to build inverted indexes or as the input of other ranking models~\cite{MalliaKTS21DeepImpact, ZhuangZ2021TILDEv2}. SPLADE-doc~\cite{FormalLPCSPLADEv2} performed document expansion in the BERT vocabulary space using the masked language modeling head of BERT. Besides, TILDE~\cite{ZhuangZ2021TILDE} utilized the query likelihood for semantic matching.

\vspace{1.0mm}
\noindent
\textbf{Term weighting.} Although classical IR models, \eg, TF-IDF~\cite{Jones2004TFIDF} and BM25~\cite{RobertsonW94BM25}, are effective for estimating the importance of terms in the document in an unsupervised manner, it does not reflect the term importance for relevant queries. DeepCT~\cite{DaiC19DeepCT} formulated term weighting as the regression problem for measuring query term recall. It learned a mapping function from contextualized word representations to term weights in a supervised manner. To mitigate the limitations of the term weighting model, \ie, vocabulary mismatch problem, recent studies~\cite{MalliaKTS21DeepImpact, ZhuangZ2021TILDEv2} adjusted term weighting scores on the expanded document. DeepImpact~\cite{MalliaKTS21DeepImpact} improved DeepCT~\cite{DaiC19DeepCT} by combining docT5query~\cite{NogueiraL19docT5query} and directly optimizing term weighting scores; it estimates the importance of terms in the document which is expanded by docT5query~\cite{NogueiraL19docT5query}. Likewise, TILDEv2~\cite{ZhuangZ2021TILDEv2} followed contextualized term weighting but employed TILDE~\cite{ZhuangZ2021TILDE} alternatively for document expansion.

\section{Proposed Model}

In this section, we propose a novel ranking model for effective and efficient first-stage retrieval, namely \emph{\textbf{Spa}rse retriever using a \textbf{D}ual document \textbf{E}ncoder (SpaDE)}, following the uni-encoder paradigm. To overcome the trade-off between effectiveness and efficiency, SpaDE adopts a dual document encoder for \emph{term weighting} and \emph{term expansion}, adjusting term importance for lexical matching and enriching additional terms for semantic matching. To train the dual encoder, we devise a \emph{co-training} strategy to collectively enjoy the strength of the dual encoder. Specifically, it consists of two stages. At the warm-up stage, each encoder is first trained independently. At the fine-tuning stage, we choose large-loss samples from each encoder, and the selected samples are used to train another encoder. Despite its simplicity, it can induce a complementary relationship by selectively training each encoder with hard samples. Furthermore, we utilize a learning-based pruning method to enforce the document-level and the corpus-level sparsity, thereby fully enjoying the efficiency of the inverted index.

In the following, we first explain the overall model architecture (Section \ref{sec:architecture}). We then introduce our key contributions, the dual document encoder (Section \ref{sec:dualencoder}), and the co-training strategy (Section \ref{sec:modeltraining},) respectively. Lastly, we explain learning-based pruning for building the efficient inverted index (Section \ref{sec:indexing}).

\subsection{Model Architecture}\label{sec:architecture}

\begin{figure}
\includegraphics[width=1.0\linewidth]{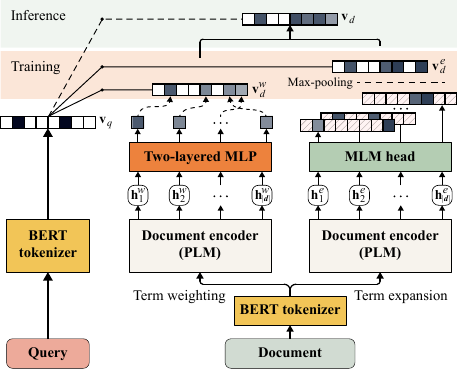}
\vskip -0.05in
\caption{Model architecture of SpaDE with the dual document encoder for term weighting and term expansion. Each encoder returns a document vector after training. For inference, two vectors are aggregated to a document vector.}\label{fig:architecture}
\vskip -0.1in
\end{figure}

Figure~\ref{fig:architecture} depicts the overall architecture of SpaDE. Our goal is to estimate the relevance function for a query $q$ and a document $d$ and return the most relevant documents. It is formulated by
\begin{equation}
sim(q, d) = g(\psi(q), \phi(d)),
\end{equation}

\noindent
where $\psi(q)$ returns a sparse query representation $\mathbf{v}_q \in \mathbb{R}^{|V|}$. Based on the BERT tokenizer, we can easily transform the query $q$ into a boolean query vector in the WordPiece vocabulary space, \eg, $|V|$ = 30,522 word-pieces. (Although we can define different vocabulary spaces depending on the tokenizer, for simplicity, we utilize the BERT tokenizer.) Note that it does not require additional costs, achieving fast query latency and reducing memory footprint to encode the query vector.

As the core part of SpaDE, $\phi(d)$ represents the dual encoder for \emph{term weighting} and \emph{term expansion}. The term weighting is responsible for adjusting the importance of terms in the document. The term expansion plays a role in generating a few relevant terms from the document and assigning their weights. After the document $d$ passes through the dual encoder, we combine two vectors into a sparse document representation $\mathbf{v}_{d} \in \mathbb{R}^{|V|}$ in the vocabulary space. During model training, sparsity is enforced by controlling both document-level and corpus-level sparsity.

Lastly, the matching function $g(\cdot, \cdot)$ is used to measure the similarity between the query vector $\mathbf{v}_q$ and the document vector $\mathbf{v}_d$. In this work, we use a simple inner product for directly optimizing the relevance score between the query and the document.

\subsection{Dual Document Encoder}\label{sec:dualencoder}

The architecture of the dual document encoder is based on the transformer encoder. Given a document $d$, we first tokenize it into sequential tokens as the input of BERT, including two special tokens [\texttt{CLS}] and [\texttt{SEP}].
\begin{equation}
\left[ \mathbf{e}_{[\texttt{CLS}]}^{w}, \mathbf{e}_{1}^{w}, \dots, \mathbf{e}_{|d|}^{w}, \mathbf{e}_{[\texttt{SEP}]}^{w} \right] \text{and} \left[ \mathbf{e}_{[\texttt{CLS}]}^{e}, \mathbf{e}_{1}^{e}, \dots, \mathbf{e}_{|d|}^{e}, \mathbf{e}_{[\texttt{SEP}]}^{e} \right],
\end{equation}

\noindent
where each embedding vector $\mathbf{e}_i^{w}$ for term weighting encoder (or $\mathbf{e}_i^{e}$ for term expansion encoder) is combined with a token embedding vector and a positional embedding vector.

Each embedding vector is then passed into the transformer encoder. The corresponding output is represented by a sequence of hidden vectors.
\begin{equation}
\left[ \mathbf{h}_{[\texttt{CLS}]}^{w}, \mathbf{h}_{1}^{w}, \dots, \mathbf{h}_{|d|}^{w}, \mathbf{h}_{[\texttt{SEP}]}^{w} \right] \text{and} \left[ \mathbf{h}_{[\texttt{CLS}]}^{e}, \mathbf{h}_{1}^{e}, \dots, \mathbf{h}_{|d|}^{e}, \mathbf{h}_{[\texttt{SEP}]}^{e} \right],
\end{equation}

\noindent
where $\mathbf{h}_{i}^{w}, \mathbf{h}_{i}^{e} \in \mathbb{R}^{m}$ for $i \in \{[\texttt{CLS}], 1, \dots, |d|, [\texttt{SEP}]\}$ and $m$ is the dimension of embedding vectors. Note that we follow the conventional structure used in BERT~\cite{DevlinCLT19BERT}.

\vspace{1mm}
\noindent
\textbf{Term weighting encoder.} As discussed in DeepCT~\cite{DaiC19DeepCT}, DeepImpact~\cite{MalliaKTS21DeepImpact} and TILDEv2~\cite{ZhuangZ2021TILDEv2}, it is effective to learn a relevance score for query-document pairs by adjusting the importance of terms in the document. Since the contextualized hidden vector for each token from PLM can capture the word's syntactic and semantic role in the local context, we compute the weighting score from hidden vectors using a two-layered MLP with the ReLU activation function. Each hidden vector $\mathbf{h}_{i}^{w}$ is projected into a weighting score $s_i$. 
\begin{equation}
s_i = f_{\text{MLP}}(\mathbf{h}_i^{w}), \ \ \text{for} \ \ i \in \{1, \dots, |d|\}.
\end{equation}

\noindent
Conceptually, the weighting score $s_i$ for each token can be represented by a one-hot vector in the vocabulary space as follows.
\begin{equation}
\mathbf{w}_i = f_{\text{one-hot}}(s_i), \ \ \text{for} \ \ i \in \{1, \dots, |d|\}.
\end{equation}

\noindent
Here, $\mathbf{w}_{i} \in \mathbb{R}^{|V|}$ is the one-hot vector for the $i$-th token. To build $\mathbf{w}_{i}$, $f_{\text{one-hot}}(\cdot)$ places $s_{i}$ into the corresponding term for $i$-th token and zeroes the rest. Finally, we aggregate them using the max pooling operation and represent the document vector $\mathbf{v}_{d}^{w} \in \mathbb{R}^{|V|}$ in the vocabulary space.
\begin{equation}
 \mathbf{v}_{d}^{w} = f_{\text{MaxPool}}(\mathbf{w}_{1}, \dots, \mathbf{w}_{|d|}).
\end{equation}

\vspace{1mm}
\noindent
\textbf{Term expansion encoder.} Since the \emph{masked language model (MLM)} head layer is pre-trained to predict a masked token, it is effective for capturing the semantic correlation between tokens. Motivated by SparTerm~\cite{BaiLWZSXWWL20SparTerm}, we compute the importance vector $\text{v}_{i} \in \mathbb{R}^{|V|}$ for each hidden vector $\mathbf{h}_i^{e}$ as follows. % $\{\text{v}_{i,j}\}_{j=1}^{|V|}$
\begin{equation}
\text{v}_{i} = \text{ReLU}(f_{\text{MLM}}(\mathbf{h}_i^{e})), \ \ \text{for} \ \ i \in \{[\texttt{CLS}], 1, \dots, |d|, [\texttt{SEP}]\},
\end{equation}

\noindent
where $f_{\text{MLM}}(\cdot)$ is the MLM head in BERT~\cite{DevlinCLT19BERT}. We represent the importance score by eliminating negative values using the ReLU activation.

Then, we utilize top-\emph{k} masking~\cite{Yang2021TopkSPLADEMask, JangKHMPYS21UHDBERT} for each $|V|$-dimensional vector $\{\text{v}_{i,j}\}_{j=1}^{|V|}$ to control the sparsity.
\begin{equation}
\label{eq:expansion_topkmasking}
\text{v}_{i,j}' = \maxj{}_k{ \left\{ \text{v}_{i,j} \right\}_{j=1}^{|V|} } \ \ \text{for} \ \ i \in \{[\texttt{CLS}], 1, \dots, |d|, [\texttt{SEP}]\},
\end{equation}
where \emph{k} is the hyperparameter to adjust the number of non-zero values in the $|V|$-dimensional vector $\text{v}_{i}$ and $\max_k$ is the operation to keep only top-\emph{k} values while others are set to zero. (Empirically, $k$ is set as 2--10.) The top-\emph{k} masking method explicitly enforces the document-level sparsity, so irrelevant or redundant tokens are removed from the document. (Refer to Section~\ref{sec:indexing} for further details.)

Finally, we aggregate them into a vocabulary-level vector by using a max-pooling operation.
\begin{equation}
\label{eq:exapnsion_maxpool}
\mathbf{v}_{d}^{e} = f_{\text{MaxPool}}(\text{v}'_{[\texttt{CLS}]}, \text{v}'_{1}, \dots, \text{v}'_{|d|}, \mathbf{v}'_{[\texttt{SEP}]}),
\end{equation}

\noindent
where $\mathbf{v}_{d}^{e} \in \mathbb{R}^{|V|}$. Several pooling operations can be used to combine multiple vectors into a single vector. The max-pooling operation shows the best performance owing to capturing the most salient values for each token, as also reported in~\cite{FormalLPCSPLADEv2}.

\vspace{1mm}
\noindent
\textbf{Aggregation.} Finally, SpaDE combines two vectors, $\mathbf{v}_{d}^{w}$ and $\mathbf{v}_{d}^{e}$, in the vocabulary space. In this process, we utilize an aggregating hyperparameter $\alpha$ to properly combine two vectors. 
\begin{equation}
\label{eq:final_score}
    \mathbf{v}_{d} = (1-\alpha)\cdot\mathbf{v}_{d}^{w} + \alpha\cdot\mathbf{v}_{d}^{e},
\end{equation}

\noindent
where $\alpha$ is the hyperparameter to balance two document vectors. (Empirically, the best performance was shown when $0.3\leq\alpha\leq0.5$.) SpaDE derives the final document vector by summing two representations of different characteristics. In other words, SpaDE (i) adjusts the importance of terms for lexical matching and (ii) endows additional terms for semantic matching.

\subsection{Co-training Strategy}\label{sec:modeltraining}

\begin{figure}[t]
\centering
\includegraphics[width=0.95\linewidth]{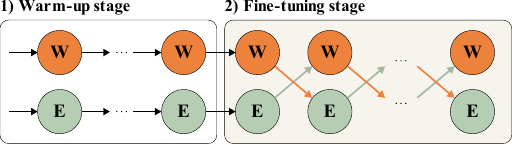}
\vskip -0.05in
\caption{Schema of our co-training strategy. Let W and E denote the term weighting encoder and the term expansion encoder, respectively. Note that a learning-based pruning method is only applied for the fine-tuning stage.}\label{fig:cotraining}
\vskip -0.1in
\end{figure}

We devise a \emph{co-training} strategy to fully enjoy the different roles of the dual encoder. The idea of co-training comes from the observation that joint learning does not achieve a clear improvement compared to training with a single encoder. (Please refer to Table~\ref{tab:ablation_component} for empirical results on the effect of training strategies.) Meanwhile, sequential training~\cite{Chen21Spar} helps improve the efficacy of training when handling two encoders. We conjecture that training with two encoders should be carefully handled, reducing the intervention from another encoder during training.

Based on the empirical observation, we design the co-training strategy with two stages (Figure~\ref{fig:cotraining}). At the warm-up stage, each encoder is first trained independently. At the fine-tuning stage, we choose \emph{large-loss} samples from each encoder and use those samples for fine-tuning another encoder. Each encoder focuses on learning different samples that their counter encoder is hard to learn, thereby compensating for the weakness of each encoder.

\vspace{1mm}
\noindent
\textbf{Warm-up stage.} We train the two encoders independently with different objective functions. For the term weighting encoder, we adopt a negative log likelihood of the positive passage. By minimizing the loss function, we can optimize the query-document relevance score as
\begin{equation}
\label{eq:weighting_loss}
    \mathcal{L}_{\text{weighting}} = -\log \frac{ \exp( \mathbf{v}_q^{\top} \mathbf{v}_{d^+}^{w})}{\exp( \mathbf{v}_q^{\top} \mathbf{v}_{d^+}^{w}) + \sum_{ d^- \in \mathcal{N}(q)} \exp( \mathbf{v}_q^{\top} \mathbf{v}_{d^-}^{w})},
\end{equation}

\noindent
which defined over a relevant query-document pair $(q, d^+) \in \mathcal{R}$ and a set of negative documents $\mathcal{N}(q)$. $\mathcal{R}$ is a training set of relevant query-document pairs.

For the term expansion encoder, we also adopt the negative log likelihood. Since SpaDE defines the relevance score using only the query terms, the term expansion encoder may not learn relevance scores for the non-query terms. To resolve the problem, we additionally introduce a point-wise loss to minimize the loss between $\mathbf{v}_{d^+}^{e}$ and $\mathbf{v}_q$; $\mathbf{v}_d^{e}$ is learned to enforce high scores for relevant query terms and low scores for irrelevant non-query terms. Note that the point-wise loss is similar to query likelihood estimation used in~\cite{ZhuangZ2021TILDE}.
\begin{equation} \label{eq:expansion_loss}
\begin{split}
    \mathcal{L}_{\text{expansion}} &= -\log \frac{ \exp( \mathbf{v}_q^{\top} \mathbf{v}_{d^+}^{\text{e}})}{\exp( \mathbf{v}_q^{\top} \mathbf{v}_{d^+}^{\text{e}}) + \sum_{ d^- \in \mathcal{N}(q)} \exp( \mathbf{v}_q^{\top} \mathbf{v}_{d^-}^{\text{e}})} 
    \\ &-\lambda\cdot\mathbf{v}_q \log \text{softmax}(\mathbf{v}_{d^+}^{e}) ,
\end{split}
\end{equation}
\noindent
where $\lambda$ is the hyperparameter to adjust the importance of the point-wise loss. (Empirically, we set $\lambda = 1$.)

\vspace{1mm}
\noindent
\textbf{Fine-tuning stage.} Given training data, one encoder first selects its large-loss samples as hard samples, and another encoder then uses those samples for fine-tuning. Specifically, each encoder extracts $\tau \%$ large-loss samples from the training set $\mathcal{R}$ for one another, \ie, $\mathcal{R}_{\text{weighting}}$ and $\mathcal{R}_{\text{expansion}}$, where $\tau$ is the hyperparameter to control the ratio of the hard training sets. To provide more informative samples for each encoder, we utilize expanded documents using docT5query~\cite{NogueiraL19docT5query}, as in DeepImpact~\cite{MalliaKTS21DeepImpact}. 
For example, when the original document is used, the term weighting encoder may pass easily handled vocabulary mismatching samples to the term expansion encoder. (Empirically, we set $\tau = 30$.)

\subsection{Learning-based Pruning}\label{sec:indexing}

Building inverted indexes using dense document vectors drops the efficiency dramatically; not only do they occupy a lot of storage space, but all query terms should traverse all documents. As pointed out in \cite{ZamaniDCLK18SNRM, BaiLWZSXWWL20SparTerm, FormalPC21SPLADE, Yang2021TopkSPLADEMask, JangKHMPYS21UHDBERT}, it is challenging to enforce sparse document representations. Previous studies exploited various techniques to obtain sparse document representations using regularization terms or parameter constraints, \eg, $L_{1}$ regularization~\cite{ZamaniDCLK18SNRM}, the FLOPS regularizer~\cite{FormalPC21SPLADE}, learning-based gating vector~\cite{BaiLWZSXWWL20SparTerm}, winner-take-all module~\cite{JangKHMPYS21UHDBERT}, and top-\emph{k} masking~\cite{Yang2021TopkSPLADEMask}.

In this paper, we adopt a learning-based pruning method to obtain sparse document representations. Specifically, SpaDE filters out some unnecessary terms to ensure the sparsity \emph{within} the document and \emph{across} the documents, \ie, \emph{document-} and \emph{corpus-level} pruning. For document-level pruning, we use top-\emph{k} masking to explicitly control sparsity by \emph{k} without complex modules. In the training process of the term expansion encoder, SpaDE leaves only top-\emph{k} values per token in the vocabulary-level vector after passing through the MLM layer as in Eq.\eqref{eq:expansion_topkmasking}. (Empirically, the average number of tokens per document is 189 when \emph{k}=10.)

Besides, we utilize corpus-level pruning based on an \emph{approximate} document frequency. Even with top-\emph{k} masking, $\mathbf{v}_d$ can still be inappropriate for building the inverted index in that it merely controls the document-level sparsity. As most of the document includes frequent terms, \eg, `\emph{the}' or `\emph{a}', it can significantly hurt the benefit of the inverted index. It is crucial to consider the corpus-level sparsity for each document, \ie, pruning terms with high document frequency. For that, SpaDE removes the terms that appear in too many documents and learns only with the remaining terms. Let $|D'|$ denote the number of documents during training, and the terms that appear more than $|D'|\times\gamma$ times in the document are removed. (Empirically, we set $\gamma$ to 0.7 and omit a detailed result due to the space limitation.) During the fine-tuning stage, we apply the document- and corpus-level pruning, ensuring the sparsity of document representations.

\section{Experimental Setup}\label{sec:setup}

\begin{table*}[h!]
\caption{Effectiveness and efficiency comparisons of various models using sparse representations on MS MARCO passage ranking. The latency is measured by averaging time over each query of the MS MARCO development set with a single thread and a single batch. Among uni-encoder models, the best model is marked \com{{\textbf{bold}}}, and the second-best model is \fcom{\ul{underlined}}. Let dT5q denote docT5query~\cite{NogueiraL19docT5query}. Results not available are denoted as ‘–’. Note that the latency is reported only for the models available for retrieval via PISA~\cite{Mallia19PISA}. Statistical significant differences ($p < 0.05$) with Bonferroni correction between the reproduced baseline models and the proposed models, \ie, SpaDE (\emph{k}=5) and SpaDE (\emph{k}=10), are reported with $*$ and $\diamond$, respectively.}

\label{tab:first_stage_marco}
\begin{center}
\vspace{-2mm}
\begin{tabular}{c|c|cc|cc|cc|cc}
\toprule
\multirow{2}{*}{Encoder} & \multirow{2}{*}{Model} & \multicolumn{2}{c|}{MS MARCO dev} & \multicolumn{2}{c|}{TREC DL 2019} & \multicolumn{2}{c|}{TREC DL 2020} & \multicolumn{1}{c}{\multirow{2}{*}{\begin{tabular}[c]{@{}c@{}}Latency\\  (ms/query)\end{tabular}}} & \multirow{2}{*}{\begin{tabular}[c]{@{}c@{}}Index size\\  (GB)\end{tabular}} \\
 &  & MRR@10 & Recall@1k & nDCG@10 & MAP & nDCG@10 & \multicolumn{1}{c|}{MAP} & \multicolumn{1}{c}{} &  \\ \midrule
- & BM25 \cite{RobertsonW94BM25} & 0.187*$^{\diamond}$ & 0.859*$^{\diamond}$ & 0.500*$^{\diamond}$ & 0.293 & 0.490*$^{\diamond}$ & 0.289 & 14 & 0.8 \\ \midrule
\multirow{7}{*}{Bi-encoder} & SparTerm~\cite{BaiLWZSXWWL20SparTerm} & 0.279 & 0.925 & - & - & - & - & - & - \\
 & UHD-BERT \cite{JangKHMPYS21UHDBERT} & 0.300 & 0.960 & - & - & - & - & - & - \\
%  & COIL-tok \cite{Gao2021COIL} & 0.341 & 0.949* & 0.660 & - & - & - & - & - \\
 & COIL-tok \cite{Gao2021COIL} & 0.341 & 0.948*$^{\diamond}$ & 0.688 & 0.457 & 0.697 & 0.452 & - & - \\
 & COIL-tok w/ dT5q \cite{Gao2021COIL} & 0.363 & 0.968 & 0.701 & 0.474 & 0.715 & 0.484 & - & - \\
 & uniCOIL \cite{LinM21uniCOIL} & 0.352 & 0.958$^{\diamond}$ & 0.702 & 0.461 & 0.673 & 0.439 & 64 & 1.4 \\
 & SPLADE-max \cite{FormalLPCSPLADEv2} & 0.340 & 0.965 & 0.682 & 0.431 & 0.671 & 0.451 & 412 & 2.0 \\
 & DistilSPLADE-max \cite{FormalLPCSPLADEv2}  & 0.369 & 0.979 & 0.728 & 0.485 & 0.710 & 0.490 & 1,606 & 4.6 \\ \midrule
\multirow{8}{*}{Uni-encoder} & DeepCT~\cite{DaiC19DeepCT} & 0.246*$^{\diamond}$ & 0.911*$^{\diamond}$ & 0.550 & 0.339 & 0.553* & 0.344 & 12 & 0.8 \\
 & docT5query \cite{NogueiraL19docT5query} & 0.276*$^{\diamond}$ & 0.946*$^{\diamond}$ & 0.641 & 0.403 & 0.617 & 0.408 & 18 & 1.2 \\
 & DeepImpact \cite{MalliaKTS21DeepImpact} & 0.327*$^{\diamond}$ & 0.948*$^{\diamond}$ & \com{\textbf{0.696}} & \com{\textbf{0.457}} & 0.652 & 0.426 & 24 & 1.6 \\
 & SPLADE-doc \cite{FormalLPCSPLADEv2}& 0.325*$^{\diamond}$ & 0.939*$^{\diamond}$ & 0.671 & 0.406 & 0.611 & 0.399 & 108 & 2.1 \\
 & TILDEv2~\cite{ZhuangZ2021TILDEv2} & 0.333*$^{\diamond}$ & 0.958$^{\diamond}$ & 0.652 & 0.408 & 0.648 & 0.432 & 31 & 2.0 \\ \cmidrule{2-10} 
 & SpaDE (\emph{k}=2) & 0.348 & 0.961 & 0.674 & 0.423 & 0.662 & 0.445 & 29 & 1.6 \\
 & SpaDE (\emph{k}=5) & \com{\textbf{0.355}} & \fcom{\ul{0.965}} & \fcom{\ul{0.682}} & \fcom{\ul{0.437}} & \com{\textbf{0.677}} & \fcom{\ul{0.453}} & 36 & 2.3 \\
 & SpaDE (\emph{k}=10) & \fcom{\ul{0.352}} & \com{\textbf{0.968}} & 0.678 & \fcom{\ul{0.437}} & \fcom{\ul{0.665}} & \com{\textbf{0.454}} & 49 & 3.6 \\ \bottomrule
\end{tabular}
\end{center}
\vspace{-2mm}
\end{table*}

\textbf{Datasets}. We mainly conduct our experiments on the MS MARCO Passage Ranking dataset~\cite{Nguyen2016msmarco}. It consists of 8.8M passages collected from Bing's results and 1M real-world queries. The average number of words in documents is 56. We use the official development set, which consists of 6,980 queries with 1.07 relevant documents per query on average. We use triples\footnote{\url{https://github.com/DI4IR/SIGIR2021}\label{deepimpact_fn}} sampled from the MS MARCO training dataset used in \cite{MalliaKTS21DeepImpact} and randomly split 3,000 queries for a validation set. We use a corpus consisting of BM25 top 100 passages to reduce a time for validation, \ie, {$|D|$ $\approx$ 300,000}. Additionally, we evaluate the models with TREC 2019 Deep Learning (DL 2019)~\cite{Craswell2020trec2019} and TREC 2020 Deep Learning (DL 2020)~\cite{Craswell2021trec2020} queries. Each dataset has 43 and 54 evaluation queries. We also verify the zero-shot performance of ranking models using BEIR~\cite{ThakurRRSG21BEIR}.

\vspace{1mm}
\noindent
\textbf{Competitive models}. We compare SpaDE with thirteen sparse representation based models, including one traditional model~\cite{RobertsonW94BM25}, seven bi-encoders~\cite{BaiLWZSXWWL20SparTerm, JangKHMPYS21UHDBERT, LinM21uniCOIL, Gao2021COIL, FormalLPCSPLADEv2}, and five uni-encoders~\cite{DaiC19DeepCT, NogueiraL19docT5query, MalliaKTS21DeepImpact, FormalLPCSPLADEv2, ZhuangZ2021TILDEv2}; we do not consider dense representation based models~\cite{KhattabZ20ColBERT,ChenHHSS21TinyBERT,Gao2021COIL,MacAvaney2020PreTTR} due to their high query inference costs. BM25~\cite{RobertsonW94BM25} is the conventional IR model using lexical matching. SparTerm~\cite{BaiLWZSXWWL20SparTerm} predicts the importance distribution in vocabulary space and represents them sparsely using a binary gating vector. UHD-BERT~\cite{JangKHMPYS21UHDBERT} learns high-dimensional sparse representations of a query and a document adopting a winner-take-all module to control sparsity. SPLADE~\cite{FormalPC21SPLADE, FormalLPCSPLADEv2} is an extension of SparTerm~\cite{BaiLWZSXWWL20SparTerm} adopting FLOPS regularizer to ensure sparsity. It has three variants, \ie, SPLADE-max, DistilSPLADE-max, and SPLADE-doc. COIL-tok~\cite{Gao2021COIL} and uniCOIL~\cite{LinM21uniCOIL} are based on COIL~\cite{Gao2021COIL} while adopting sparse representations. DeepCT~\cite{DaiC19DeepCT} leverages BERT~\cite{DevlinCLT19BERT} to adjust term frequencies in the document. Next, docT5query~\cite{NogueiraL19docT5query} extends the document terms by generating relevant queries from the documents using T5~\cite{Raffel2020t5}. DeepImpact~\cite{MalliaKTS21DeepImpact} is an improved term weighting model of DeepCT~\cite{DaiC19DeepCT} to directly learn the term scores by adopting docT5query~\cite{NogueiraL19docT5query} for document expansion. Besides, TILDEv2~\cite{ZhuangZ2021TILDEv2} is a term weighting model that improves TILDE~\cite{ZhuangZ2021TILDE}.

\vspace{1mm}
\noindent
\textbf{Reproducibility}. We implemented SpaDE using PyTorch. The pre-trained language model of SpaDE was initialized with BERT$_{\text{base}}$. We used the BERT WordPiece vocabulary ($|V|$ = 30,522) and set $m$ to 768. We used the Adam optimizer with a learning rate of 1e-5 and 5e-6 for term weighting and term expansion, respectively. We set the max sequence length of BERT to 256, dropout rate to 0.1, and batch size to 32. We conducted a grid search for $\lambda$ among \{0.1, 1, 10\} and $\tau$ in [0, 100] with the step size 10 and set to 1 and 30, respectively. For an aggregating hyperparameter $\alpha$, we searched among [0,1] with the step size 0.1 on the valid set for each run. We set the warm-up stage to 32,000 iterations. We conducted all the optimizations on the valid set and kept the best checkpoint using MRR@10 by evaluating every 1,600 iterations; we stopped training after MRR@10 had not been improved five times, which stops after 54,400 iterations on average. For the document-level pruning, top-$k$ masking is 2--10. For corpus-level pruning, we set $\gamma$=0.7 after tuning in [0.1, 0.9]. For the MS MARCO passage set, we used expanded documents\footref{deepimpact_fn} by docT5query~\cite{NogueiraL19docT5query}. We quantized the token scores into eight bits for SpaDE. For all models, we built the inverted index by Anserini~\cite{Yang0L17anserini} and exported it to the Common Index File Format~\cite{Lin20CIFF} before being imported into PISA~\cite{Mallia19PISA} following \cite{Mackenzie21WackyWeight}. We adopted in-batch negatives where all passages for other queries in the same batch are considered negative. Experimental results for SpaDE are averaged over three runs with different seeds. For BM25~\cite{RobertsonW94BM25}, we followed the official guide\footnote{\url{http://anserini.io/}}. For uniCOIL\footnote{\url{https://github.com/luyug/COIL/tree/main/uniCOIL}\label{coil_url}}~\cite{LinM21uniCOIL}, COIL-tok\footref{coil_url}~\cite{Gao2021COIL}, SPLADE-max\footnote{\url{https://github.com/naver/splade}}~\cite{FormalLPCSPLADEv2}, DistilSPLADE-max\footnote{\url{https://github.com/castorini/pyserini/blob/master/docs/experiments-spladev2.md}}~\cite{FormalLPCSPLADEv2}, DeepCT\footnote{\url{https://github.com/AdeDZY/DeepCT}}~\cite{DaiC19DeepCT}, DeepImpact\footref{deepimpact_fn}~\cite{MalliaKTS21DeepImpact}, and TILDEv2\footnote{\url{https://github.com/ielab/TILDE}}~\cite{ZhuangZ2021TILDEv2}, we used the official code provided by the authors. For docT5query~\cite{NogueiraL19docT5query}, we used the published predicted queries and added 40 predictions to each passage as recommended. For COIL-tok~\cite{Gao2021COIL}, we also reported the results combined with docT5query~\cite{NogueiraL19docT5query} expansion fair comparison. We do not report the efficiency for COIL-tok~\cite{Gao2021COIL} because it is unable to compare in the same environment, \eg, PISA~\cite{Mallia19PISA}. For SPLADE-doc~\cite{FormalLPCSPLADEv2}, we reproduced the model following the hyperparameters from the paper~\cite{FormalLPCSPLADEv2}. For SparTerm~\cite{BaiLWZSXWWL20SparTerm} and UHD-BERT~\cite{JangKHMPYS21UHDBERT}, the results are obtained from the original paper. We conducted all experiments on a desktop with 2 NVidia GeForce RTX 3090, 512 GB memory, and a single Intel Xeon Gold 6226. 
All the source code is available\footnote{\url{https://github.com/eunseongc/SpaDE}}.

\vspace{1mm}
\noindent
\textbf{Evaluation metrics}. To measure the effectiveness, we use recall, mean reciprocal rank (MRR), normalized discounted cumulative gain (nDCG), and mean average precision (MAP) with retrieval size $K$. Recall is defined as $\frac{\sum_{i=1}^N{rel_i}}{k}$, where $i$ is the position in the list, $k$ is the number of relevant documents and $rel_i \in \{0, 1\}$ indicates whether the $i$-th document is relevant to the query or not. We report Recall for $K$=1000. MRR is defined as $\frac{1}{|Q|}\sum_{i=1}^{|Q|}{\frac{1}{rank_i}}$, where $rank_i$ refers to the rank position of the first relevant document for the $i$-th query. nDCG considers the order of retrieved documents in the list. DCG@K is defined as $\sum_{i=1}^{K}\frac{2^{rel_i}-1}{log_2{(i+1)}}$ where $rel_i$ is the graded relevance of the result at position $i$. nDCG is the ratio of DCG to the maximum possible DCG for the query, which occurs when the retrieved documents are presented in decreasing order of relevance. MRR and nDCG is the official metric of MS MARCO Passage Ranking and TREC Deep Learning Track, respectively. We also report MAP used in existing studies~\cite{MalliaKTS21DeepImpact, KhattabZ20ColBERT}. For MRR and nDCG, we set $K$=10. To measure the efficiency, we report the average latency time for processing each query with a single thread and a single batch.
\section{Results and Analysis}\label{sec:results}
We evaluate the effectiveness and efficiency of SpaDE with competing models and summarize meaningful results.

\begin{itemize}[leftmargin=5mm]
\item SpaDE achieves state-of-the-art performance with acceptable latency. On the MS MARCO development set, SpaDE achieves MRR@10 and Recall@1k of 0.352 and 0.968, outperforming the best competitive uni-encoder model by 6.67\% and 1.04\% with fast query inference time. (Section~\ref{sec:exp_Performance})

\item SpaDE proves generalization capabilities in the zero-shot setting. For search tasks from the BEIR dataset, SpaDE shows nDCG@10 of 0.462 on average, outperforming the best competing model by 3.13\%. (Section~\ref{sec:exp_Performance})

\item SpaDE addresses the trade-off between effectiveness and efficiency by showing higher accuracy than other baselines with comparable latency. (Section~\ref{sec:exp_in-depth})

\item The co-training strategy for the dual document encoder shows better accuracy than simple joint learning by 15.4\% in MRR@10. (Section~\ref{sec:exp_ablation})

\item The learning-based pruning method improves efficiency by 3.4x faster in query latency without sacrificing the retrieval effectiveness. (Section~\ref{sec:exp_ablation})

\end{itemize}


\begin{table}[t!] \small
\caption{nDCG@10 of uni-encoder models on BEIR~\cite{ThakurRRSG21BEIR}. The best model is marked {\com{\textbf{bold}}}, and the second-best model is {\fcom{\underline{underlined}}}. Let dT5q and S-doc denote docT5query and SPLADE-doc. For fair evaluation, we exclude expanded terms using docT5query in SpaDE (\emph{k}=10).}
\label{tab:beir_result}
\vspace{-0.2mm}
\begin{center}
\begin{tabular}{cccccc}
\toprule
\multicolumn{1}{c|}{Corpus} & BM25 & DeepCT & dT5q & S-doc & SpaDE \\ \midrule
\multicolumn{6}{c}{BEIR Search Tasks} \\ \midrule
\multicolumn{1}{c|}{DBPedia~\cite{HasibiNXBBKC17DBPedia}} & 0.273 & 0.177 & 0.331 & \fcom{\ul {0.338}} & \com{\textbf{0.353}} \\
\multicolumn{1}{c|}{FiQA-2018~\cite{MaiaHFDMZB18FiQA2018}} & 0.236 & 0.191 & \com{\textbf{0.336}} & 0.233 & \fcom{\ul {0.288}} \\
\multicolumn{1}{c|}{HotpotQA~\cite{Yang0ZBCSM18HotpotQA}} & \fcom{\ul {0.603}} & 0.503 & 0.580 & 0.537 & \com{\textbf{0.629}} \\
\multicolumn{1}{c|}{NFCorpus~\cite{BotevaGSR16NFCorpus}} & 0.325 & 0.283 & \com{\textbf{0.328}} & 0.310 & \fcom{\ul{0.327}} \\
\multicolumn{1}{c|}{NQ~\cite{KwiatkowskiPRCP19NQ}} & 0.329 & 0.188 & 0.399 & \fcom{\ul{0.400}} & \com{\textbf{0.476}} \\
\multicolumn{1}{c|}{TREC-COVID~\cite{VoorheesABDHLRS20TRECCOVID}} & 0.656 & 0.406 & \com{\textbf{0.713}} & 0.568 & \fcom{\ul{0.700}} \\ \midrule
\multicolumn{1}{c|}{Average} & 0.404 & 0.291 & \fcom{\ul{0.448}} & 0.398 & \com{\textbf{0.462}} \\  \midrule
\multicolumn{6}{c}{BEIR Semantic Relatedness Tasks} \\ \midrule
\multicolumn{1}{c|}{Touché(v2)~\cite{BondarenkoGFBAP21aTouchev2}} & \com{\textbf{0.367}} & 0.156 & \fcom{\ul{0.347}} & 0.241 & 0.276 \\ 
\multicolumn{1}{c|}{ArguAna~\cite{Wachsmuth18Arguana}} & \fcom{\ul {0.315}} & 0.309 & \com{\textbf{0.349}} & 0.230 & 0.288 \\
\multicolumn{1}{c|}{Climate-FEVER~\cite{Diggelmann2020CLIMATE-FEVER}} & \com{\textbf{0.213}} & 0.066 & \fcom{\ul {0.201}} & 0.091 & 0.167 \\
\multicolumn{1}{c|}{FEVER~\cite{Thorne2018FEVER}} & \com{\textbf{0.753}} & 0.353 & \fcom{\ul {0.714}} & 0.627 & 0.691 \\
\multicolumn{1}{c|}{SCIDOCS~\cite{Cohan2020SCIDOCS}} & \fcom{\ul {0.158}} & 0.124 & \com{\textbf{0.162}} & 0.137 & 0.149 \\
\multicolumn{1}{c|}{SciFact~\cite{David2020SciFact}} & 0.665 & 0.630 & \fcom{\ul{0.675}} & 0.649 & \com{\textbf{0.676}} \\ \midrule
\multicolumn{1}{c|}{Average} & \com{\textbf{0.412}} & 0.273 & \fcom{\ul{0.408}} & 0.329 &  0.374 \\ \bottomrule
\end{tabular}
\end{center}
\end{table}

\subsection{Effectiveness vs. Efficiency}\label{sec:exp_Performance}

\noindent\textbf{Full-ranking evaluation on MS MARCO.} Table \ref{tab:first_stage_marco} reports the first-stage retrieval accuracy on the MS MARCO passage ranking dataset. The key observations are as follows: (i) SpaDE shows the best performance among all uni-encoder models and is comparable to bi-encoder models. It is well-suited as a first-stage retriever in terms of high recall and low computational cost. In particular, when expanding two terms per token, it achieves higher recall with lower query latency than other models except for models using BM25 algorithms, \eg, BM25~\cite{RobertsonW94BM25}, DeepCT~\cite{DaiC19DeepCT}, and docT5query~\cite{NogueiraL19docT5query}. (ii) Other than uniCOIL~\cite{LinM21uniCOIL}, bi-encoder models took much more query processing time than uni-encoder models since they expand query terms, increasing the number of matching terms. Although SpaDE belongs to the uni-encoder models, it achieves comparable retrieval effectiveness to bi-encoders with much lower query latency. Query encoding time is excluded for bi-encoder models in measuring latency. Note that uniCOIL~\cite{LinM21uniCOIL} squeezes the dimension of a token embedding of COIL-tok~\cite{Gao2021COIL} to one and is always more efficient or similar. (iii) Increasing the expanded number of terms per token $k$ improves the recall, which indicates that the term expanding encoder effectively produces essential terms that are not covered by the term weighting encoder.

In MS MARCO development queries, SpaDE achieves the best performance among the uni-encoder models showing a clear improvement, \eg, $+$0.022 in MRR@10. This gain is 3.6x bigger than the improvement between the best and the second-best competitive models. TREC DL 2019 and 2020 can be considered closer to the real-world scenarios since the average number of relevant documents per query is 58.2 and 30.9, and the relevance score is judged on a four-point scale. As reported in Table~\ref{tab:first_stage_marco}, SpaDE mostly shows better performance than all uni-encoders on TREC DL 2019 and 2020, except for DeepImpact~\cite{MalliaKTS21DeepImpact}.

\vspace{1mm}
\noindent\textbf{Zero-shot evaluation on BEIR.} Table \ref{tab:beir_result} reports the zero-shot performance of the first-stage retrieval models on the BEIR~\cite{ThakurRRSG21BEIR} dataset. Among uni-encoder models, the models that adopt the document expansion are excluded from comparison if the results are not available on the BEIR leaderboard\footnote{\url{https://github.com/beir-cellar/beir}}. We divide the datasets into two categories by task types in ~\cite{ThakurRRSG21BEIR}; search tasks (\ie, bio-medical information retrieval and question answering) and semantic relatedness tasks (\ie, argument retrieval, citation-relatedness, and claim verification) considering the nature of queries following the existing work~\cite{SanthanamKSPZ21ColBERTv2}. It is found that SpaDE shows competitive performance for search tasks, which is similar to passage retrieval. SpaDE shows higher average accuracy than the other baselines, implying it generalizes well to diverse datasets. For semantic relatedness tasks, the models using term frequency scores, \eg, BM25~\cite{RobertsonW94BM25} and docT5query~\cite{NogueiraL19docT5query}, show better results. Compared to the search task, SpaDE is less effective for the semantic relatedness task, but it still shows better generalization capabilities compared to other PLM-based models, \ie, DeepCT~\cite{DaiC19DeepCT} and SPLADE-doc~\cite{FormalLPCSPLADEv2}.

\begin{table}[t!]\small
\caption{Effectiveness of the uni-encoder models on the subsets of the MS MARCO development set, considering the degree of query and document overlapping. The best model is marked {\com{\textbf{bold}}}, and the second-best model is {\fcom{\underline{underlined}}}. The number in parentheses indicates the number of queries. For SpaDE, significant differences ($p < 0.05$) with Bonferroni correction are reported with $^{\diamond}$.}
\label{tab:first_stage_nomatch}
\vspace{-0.2mm}
\begin{tabular}{c|c|ccc}
\toprule
Dataset & Model & \multicolumn{1}{c}{MRR@10} & \multicolumn{1}{c}{Recall@1k} & \multicolumn{1}{c}{MAP} \\ \midrule
\multirow{8}{*}{\begin{tabular}[c]{@{}c@{}}Match$_{\le1}$\\ (291)\end{tabular}}& BM25~\cite{RobertsonW94BM25} & 0.016$^{\diamond}$ & 0.397$^{\diamond}$ & 0.012$^{\diamond}$ \\
 & DeepCT~\cite{DaiC19DeepCT}& 0.034$^{\diamond}$ & 0.604$^{\diamond}$ & 0.029$^{\diamond}$ \\
 & docT5query~\cite{NogueiraL19docT5query} & 0.106 & 0.785 & 0.102 \\
 & DeepImpact~\cite{MalliaKTS21DeepImpact} & 0.135 & 0.721 & 0.128 \\
 & SPLADE-doc~\cite{FormalLPCSPLADEv2} & 0.130 & 0.706$^{\diamond}$ & 0.124 \\
 & TILDEv2~\cite{ZhuangZ2021TILDEv2} & \fcom{\ul{0.144}} & \fcom{\ul{0.793}} & \com{\textbf{0.140}} \\ \cmidrule{2-5} 
 & SpaDE (\emph{k}=10)  & \com{\textbf{0.145}} & \com{\textbf{0.799}} & \fcom{\ul{0.137}} \\ \midrule
\multirow{8}{*}{\begin{tabular}[c]{@{}c@{}}Match$_{>1}$\\ (5,959)\end{tabular}}& BM25~\cite{RobertsonW94BM25} & 0.154$^{\diamond}$ & 0.847$^{\diamond}$ & 0.150$^{\diamond}$ \\
 & DeepCT~\cite{DaiC19DeepCT} & 0.223$^{\diamond}$ & 0.915$^{\diamond}$ & 0.218$^{\diamond}$ \\
 & docT5query~\cite{NogueiraL19docT5query} & 0.261$^{\diamond}$ & 0.948$^{\diamond}$ & 0.257$^{\diamond}$ \\
 & DeepImpact~\cite{MalliaKTS21DeepImpact} & 0.311$^{\diamond}$ & 0.953$^{\diamond}$ & 0.306$^{\diamond}$ \\
 & SPLADE-doc~\cite{FormalLPCSPLADEv2} & 0.305$^{\diamond}$ & 0.943$^{\diamond}$ & 0.299$^{\diamond}$ \\
 & TILDEv2~\cite{ZhuangZ2021TILDEv2} & \fcom{\ul{0.313}}$^{\diamond}$ & \fcom{\ul{0.961}}$^{\diamond}$ & \fcom{\ul{0.307}}$^{\diamond}$ \\ \cmidrule{2-5} 
 &  SpaDE (\emph{k}=10)  & \com{\textbf{0.337}} & \com{\textbf{0.973}} & \com{\textbf{0.332}} \\ \midrule
\multirow{8}{*}{\begin{tabular}[c]{@{}c@{}}Match$_{\text{all}}$\\ (730)\end{tabular}} & BM25~\cite{RobertsonW94BM25}& 0.449$^{\diamond}$ & 0.984$^{\diamond}$ & 0.442$^{\diamond}$ \\
 & DeepCT~\cite{DaiC19DeepCT} & 0.516 & \fcom{\ul{0.996}} & 0.507 \\
 & docT5query~\cite{NogueiraL19docT5query} & 0.461$^{\diamond}$ & \com{\textbf{0.997}} & 0.454$^{\diamond}$ \\
 & DeepImpact~\cite{MalliaKTS21DeepImpact} & 0.535 & \fcom{\ul{0.996}} & 0.528 \\
 & SPLADE-doc~\cite{FormalLPCSPLADEv2} & \fcom{\ul{0.560}} & \com{\textbf{0.997}} & \fcom{\ul{0.553}} \\
 & TILDEv2~\cite{ZhuangZ2021TILDEv2} & \com{\textbf{0.565}} & \com{\textbf{0.997}} & \com{\textbf{0.556}} \\ \cmidrule{2-5} 
 &  SpaDE (\emph{k}=10)   & 0.557 & \com{\textbf{0.997}} & 0.550 \\ \bottomrule
\end{tabular}
\end{table}

\vspace{1mm}
\noindent\textbf{Break-down analysis on MS MARCO.} Table \ref{tab:first_stage_nomatch} shows a break-down analysis of the first-stage retrieval accuracy on the MS MARCO development set over uni-encoders. Depending on the number of overlapping terms between the queries and the documents, we divided the queries into three subsets: Match$_{\le1}$, Match$_{>1}$, and  Match$_{\text{all}}$. A query is classified as Match$_{\text{all}}$ if all query terms are matched with the relevant document in the WordPiece vocabulary. When the number of query terms matched with the document is zero (23 queries) or one (268 queries), the query is classified into a Match$_{\le1}$. The rest of the queries are classified as Match$_{>1}$. The query set may vary depending on the vocabulary set, but we judged that the performance comparison is reasonable based on the performance of Match$_{\text{all}}$.

The key observations are as follows: (i) SpaDE shows competitive performance with the other baselines, especially in Match$_{\le1}$ and Match$_{>1}$, and remarkably outperforms the others in most of the metrics. It implies that SpaDE expands meaningful terms to the document by effectively using the dual encoder. (ii) By comparing DeepCT~\cite{DaiC19DeepCT} and docT5query~\cite{NogueiraL19docT5query}, we can observe that term expanding is essential for semantic matching (Match$_{\le1}$ and Match$_{>1}$) and term weighting is important for effective lexical matching (Match$_{\text{all}}$). (iii) The models considering both term weighting and term expansion, \ie, SpaDE, DeepImpact~\cite{MalliaKTS21DeepImpact}, SPLADE-doc~\cite{FormalLPCSPLADEv2}, and TILDEv2~\cite{ZhuangZ2021TILDEv2}, show effective performance for all query sets, proving that adopting both solutions can be effective.

\vspace{1mm}
\begin{table}[t]\small
\caption{MRR@10 of re-ranking results using MiniLM~\cite{WangBHDW21minilmv2} on the MS MARCO development set over various depths. The best model is marked \com{\textbf{bold}}.} 
\label{tab:re-rank}
\begin{center}
\centering
\begin{tabular}{c|ccc|c}
\toprule
Depth       & BM25  & DeepImpact     & TILDEv2        & SpaDE (\emph{k}=10)  \\ \midrule
First-stage & 0.187 & 0.327          & 0.333          & \com{\textbf{0.352}} \\ \midrule
10          & 0.285 & 0.378          & 0.383          & \com{\textbf{0.391}} \\
20          & 0.322 & 0.391          & 0.396          & \com{\textbf{0.400}} \\
50          & 0.351 & 0.398          & 0.403          & \com{\textbf{0.405}} \\
100         & 0.367 & 0.402          & 0.404          & \com{\textbf{0.405}} \\
\bottomrule
\end{tabular}
\end{center}
\end{table}

\noindent\textbf{Re-ranking evaluation.} Table~\ref{tab:re-rank} shows the re-ranking performance on the MS MARCO development set over various depths with MiniLM~\cite{WangBHDW21minilmv2} as a cross-encoder re-ranker. Although the original re-ranking task utilizes 1000 passages, it is still a high cost for cross-encoder models. Therefore, in the real-world scenario, re-ranking with fewer passages can be more practical. It is attractive that SpaDE shows outstanding performance in shallow depths, \eg, 0.391 of MRR@10 in the depth of 10, compared to other baselines.

\subsection{In-depth Analysis}\label{sec:exp_in-depth}

Figure~\ref{fig:expand} depicts the performance of SpaDE varying the number of expanded terms per token $k$ with other baselines. (i) As $k$ increases, SpaDE improves the accuracy, and its performance tends to converge when $k$=10. Considering the trade-off between effectiveness and efficiency, it seems appropriate to expand ten terms per token, leading to 189 terms per document on average. (ii) By varying $k$, SpaDE achieves higher efficiency than other baselines with comparable effectiveness. Specifically, its latency time is much faster than uniCOIL~\cite{LinM21uniCOIL} when $k$=5 while achieving better performance even without a complex query encoder.

\begin{table}[t!] \small
\caption{Ablation study of SpaDE on the MS MARCO development set in training the dual encoder. Let dT5q denote docT5query~\cite{NogueiraL19docT5query}.}
\label{tab:ablation_component}
\begin{tabular}{c|ccc|c}
\toprule
Model & MRR@10 & Recall@1k & MAP & Latency \\ \midrule
SpaDE (\emph{k}=10) & \textbf{0.352} & \textbf{0.968} & \textbf{0.347} & 49 \\ \midrule
Term weighting only & 0.309 & 0.952 & 0.307 & 16 \\
Term expansion only & 0.316 & 0.950 & 0.309 & 42 \\ \midrule
Joint learning & 0.305 & 0.950 & 0.300 & 56 \\
w/o co-training & 0.342 & 0.965 & 0.336 & 48 \\
w/o dT5q expansion & 0.344 & 0.958 & 0.339 & 45 \\
w/o corpus-level pruning  & 0.352 & 0.965 & 0.336 & 165 \\  \bottomrule
\end{tabular}
\end{table}

\subsection{Ablation Study}\label{sec:exp_ablation}

Table~\ref{tab:ablation_component} shows the effectiveness of SpaDE with various training methods. (i) The proposed co-training strategy remarkably improves the accuracy compared to using a single encoder by more than 11.4\% in MRR@10. Using a dual encoder always shows better performance than using a mere single encoder. It implies that each encoder successfully captures complementary information to another. However, when we jointly train the dual encoder, each encoder fails to perform well due to unnecessary intervention during training. Instead of minimizing Eq.~\eqref{eq:weighting_loss} and Eq.~\eqref{eq:expansion_loss}, joint training minimizes $\mathcal{L}_{\text{joint}}$ which replaces $\mathbf{v}_{d}^{e}$ in Eq.~\eqref{eq:expansion_loss} to $\mathbf{v}_{d}$. We fix $\alpha$ to 0.3 to aggregate two vectors in joint training, and it directly learns combined document representations. (ii) When we do not use the co-training strategy, MRR@10 drops from 0.352 to 0.342, indicating that the co-training strategy is effective in accuracy gains. Note that we apply the pruning method from the start of training in this setting. (iii) The term expansion component can alleviate the vocabulary mismatch problem without expanded documents beforehand using docT5query~\cite{NogueiraL19docT5query}. (iv) Without the corpus-level pruning, query latency is increased by 3.4x while there is no gain in performance. Empirically, there is a trade-off between efficiency and effectiveness over varying cutoff ratios $\gamma$, and the performance seems to be converged when $\gamma$=0.7 at rapid query latency. We set the cutoff ratio $\gamma$ to 0.7 which leads to 60 stopwords, \eg, $\{$`\emph{the}', `\emph{it}', `\emph{what}', `\emph{for}', `\emph{as}', ...$\}$, to be removed on average.

\section{Conclusion}\label{sec:conclusion}

In this paper, we proposed a novel uni-encoder model, \emph{\textbf{Spa}rse retriever using a \textbf{D}ual document \textbf{E}ncoder (SpaDE)}, to alleviate the trade-off between effectiveness and efficiency of the IR system. We adopted a dual document encoder for lexical and semantic matching and developed a co-training strategy to mitigate the training intervention between encoders. We also utilized document- and corpus-level pruning during model training, enabling efficient retrieval using the inverted index. Experimental results showed that SpaDE achieves state-of-the-art performance among uni-encoder models with acceptable query latency, notably preferable for commercial IR systems.

\section*{Acknowledgments} 
This work was supported by Naver Corporation. Also, it was supported by Institute of Information \& communications Technology Planning \& Evaluation (IITP) grant funded by the Korea government (MSIT) (No.2019-0-00421 AI Graduate School Support Program (SKKU) and 2022-0-00680), and the National Research Foundation of Korea (NRF) grant funded by the Korea government (MSIT) (NRF-2018R1A5A1060031).

\newpage
\bibliographystyle{ACM-Reference-Format}
\bibliography{references}

\newpage

\appendix

\begin{table}[t!]\small
\caption{Effectiveness of various models on MS MARCO document ranking. Among models using sparse representations, the best model is marked \com{{\textbf{bold}}}. Let DR and dT5q denote Dense Retrieval and docT5query~\cite{NogueiraL19docT5query}. Results not available are denoted as ‘–’.}

\label{tab:first_stage_document_2}

\centering
\begin{tabular}{c|c|c|cc}
\toprule
Rep.          & Encoder                      & Model        & MRR@100 & R@100 \\ \midrule
\multirow{5}{*}{Dense}  & \multirow{5}{*}{Bi}          & DR (Rand Neg)~\cite{Zhan2021ADORE}    & 0.330   & 0.859      \\
                        &                              & DR (BM25 Neg)~\cite{Zhan2021ADORE}         & 0.316   & 0.794      \\
                        &                              & ANCE~\cite{XiongXLTLBAO21ANCE}         & 0.377   & 0.894      \\
                        &                              & ADORE~\cite{Zhan2021ADORE}        & 0.405   & 0.919      \\
                        &                              & RepCONC~\cite{Zhan22RepCONC}      & 0.399   & 0.911      \\ \midrule
\multirow{7}{*}{Sparse} & \multirow{3}{*}{Bi}  & COIL~\cite{Gao2021COIL}         & -       & -          \\
                        &                              & uniCOIL~\cite{LinM21uniCOIL}      & 0.341   & 0.864      \\
                        &                              & uniCOIL-dT5q~\cite{LinM21uniCOIL} & 0.353   & 0.886      \\ \cmidrule{2-5}
                        & \multirow{1}{*}{-}           & BM25~\cite{RobertsonW94BM25}         & 0.278   & 0.807      \\ \cmidrule{2-5}
                        & \multirow{3}{*}{Uni} & HDCT~\cite{DaiC20HDCT}         & 0.320   & 0.843      \\ 
                        &                              & docT5query~\cite{NogueiraL19docT5query}   & 0.327   & 0.861      \\
                        &                              & SpaDE (\emph{k}=5) & \com{\textbf{0.369}}   & \com{\textbf{0.899}}    \\ \bottomrule
\end{tabular}
\end{table}

\section{Document Retrieval}\label{sec:app_first_stage_document}

Table~\ref{tab:first_stage_document_2} shows the performance of SpaDE with other baselines on the MS MARCO document ranking dataset. DR (Random Neg) and DR (BM25 Neg) represent Dense Retrieval models trained with random and BM25 negatives, respectively. All dense retrieval models use RoBERTa-base~\cite{Liu2019RoBERTa} as an encoder and \texttt{[CLS]} token embedding for query and document representations. The experimental results of them are copied from ~\cite{Zhan2021ADORE, Zhan22RepCONC}. uniCOIL and uniCOIL-dT5q performed indexing on segmented passages, and the score of the segmented passage that obtained the highest score is used as the score of the document, \ie, MaxP technique~\cite{Dai2019bertmaxp}. Other models, including SpaDE, use only the first part of the document truncated to BERT's max length. We use official metrics for MS MARCO document ranking task, \ie, MRR@100 and Recall@100. As a result, SpaDE achieves the best performance among baselines using sparse representations. It may seem similar to the passage ranking task results, but spade's effectiveness is more highlighted in long documents. For example, in the passage ranking, MRR@10 performance of uniCOIL and SpaDE were 0.351 and 0.353, respectively, but in the document ranking, SpaDE significantly outperforms it by 0.369 versus 0.353. Secondly, ANCE shows high ranking performance, \ie, MRR, compared to SpaDE, but SpaDE has better recall performance. If compared considering the same MRR performance, uni-encoder outperforms bi-encoder in terms of Recall. This trend, which was not shown before, suggests that focusing on intrinsic lexical matching signals may be more effective than learning representations of both documents and queries in the long document environment where vocabulary mismatching occurs relatively less. In other words, as the first-stage retriever in document ranking, the uni-encoder method may be more favorable than the bi-encoder method.

\begin{figure}[t]
\centering
\includegraphics[width=1.0\linewidth]{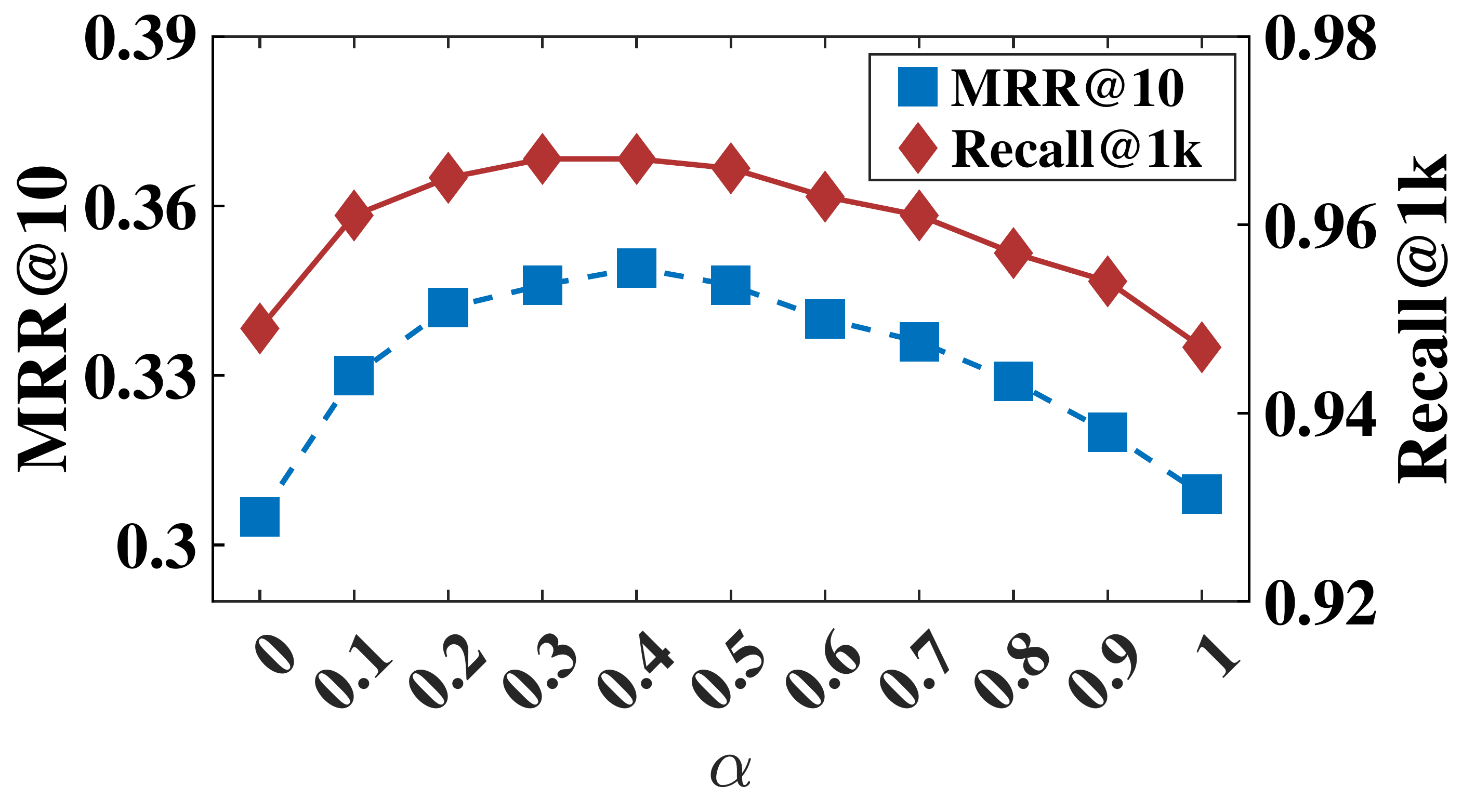}
\caption{MRR@10 and Recall@1k of SpaDE (\emph{k}=10) over varying $\alpha$. Note that we chose $\alpha$ based on the valid set.}\label{fig:alpha}
\vskip -0.1in
\end{figure}
\begin{figure}[t]
\centering
\includegraphics[width=1.0\linewidth]{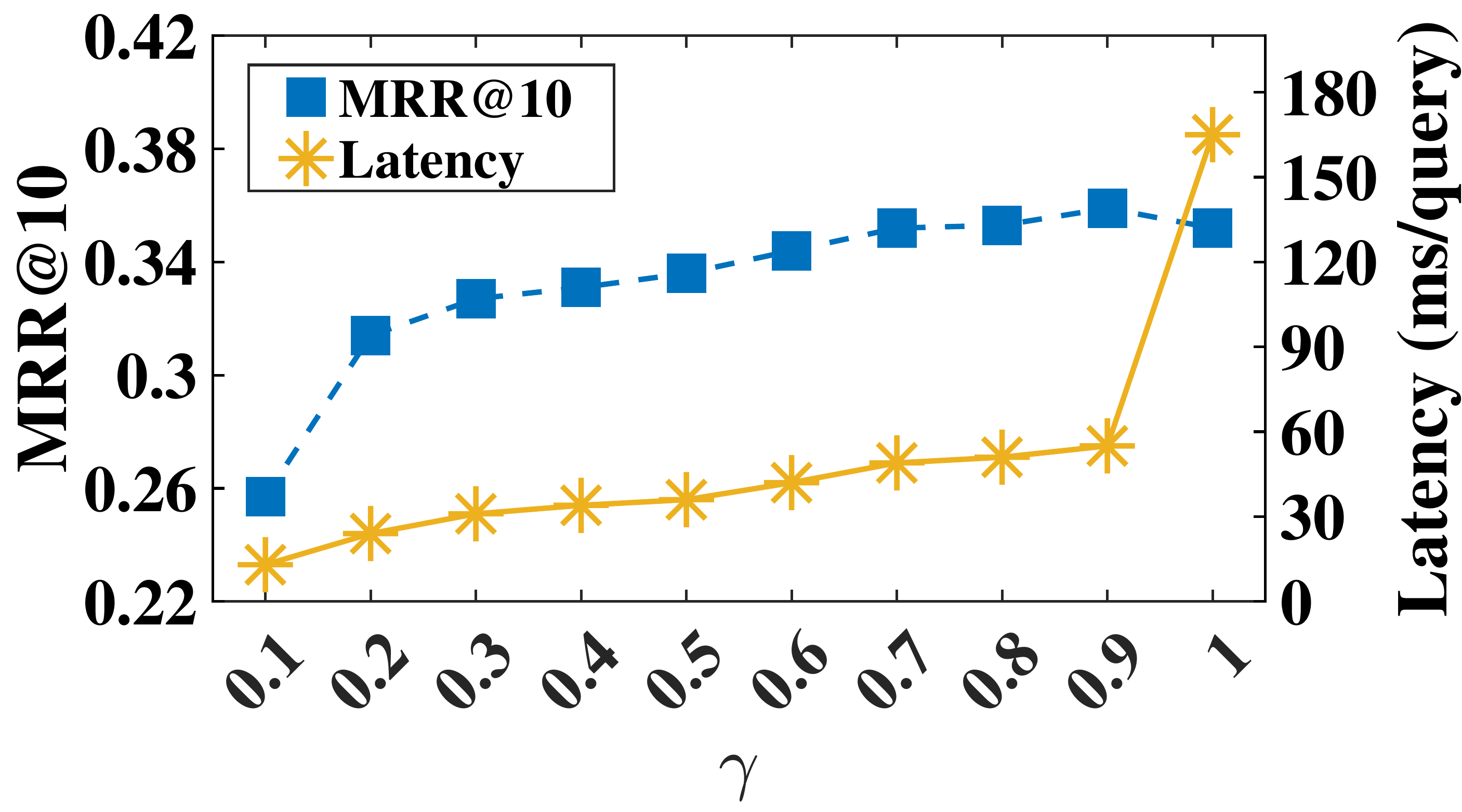}
\vskip -0.1in
\caption{MRR@10 and query latency (in ms) of SpaDE (\emph{k}=10) over varying $\gamma$ of corpus-level pruning. When $\gamma$=0.7, we prune terms appearing in over 70\% of documents while training. The latency is measured with PISA~\cite{Mallia19PISA} using Block-Max WAND~\cite{DingS11blockmaxwand}.}\label{fig:pruning}
\end{figure}

\section{Effects of $\alpha$}\label{sec:alpha}

Figure~\ref{fig:alpha} shows the effect of the aggregating hyperparameter $\alpha$. Using the dual encoder, \ie, $0 < \alpha < 1$, mostly shows better performance than using a mere single encoder, \ie, $\alpha=0$ or $\alpha=1$, depicting that each encoder only captures complementary information to another. SpaDE (\emph{k}=10) shows the best performance at $\alpha = 0.4$ with 0.352 in MRR@10, implying that the term weighting encoder is more dominant than the term expansion encoder.

\section{Effects of corpus-level pruning}\label{sec:pruning}
Figure~\ref{fig:pruning} shows the effect of corpus-level pruning of SpaDE. Since new relevant terms are appended through the term expansion component without cutoff based on an approximate document frequency, the number of elements in the term-document matrix is enormous, at about 1.1 billion, implying query processing is very costly, \eg, 510 ms per query. To reduce the query latency, we use corpus-level pruning introduced in Section~\ref{sec:indexing}. Specifically, during the model training, terms with high document frequency are pruned, and the model learns document representations only with the remaining terms. There is a trade-off between effectiveness and efficiency depending on the cutoff ratio. When the threshold is set from 1.0 to 0.2, MRR@10 drops from 0.352 to 0.314, while the inference is about 6.9x faster. Also, it shows the highest performance in MRR@10 when $\gamma$=0.9, implying that it can be helpful for retrieval effectiveness to learn document representations, excluding redundant terms. We set the cutoff ratio $\gamma$ to 0.7 which leads to 60 stop words, \eg, $\{$`\emph{the}', `\emph{type}', `\emph{it}', `\emph{what}', `\emph{for}', `\emph{as}', ...$\}$,  to be removed on average and the average number of tokens per document is 189.

\begin{figure}
\centering
\begin{tabular}{cc}
\includegraphics[width=0.22\textwidth]{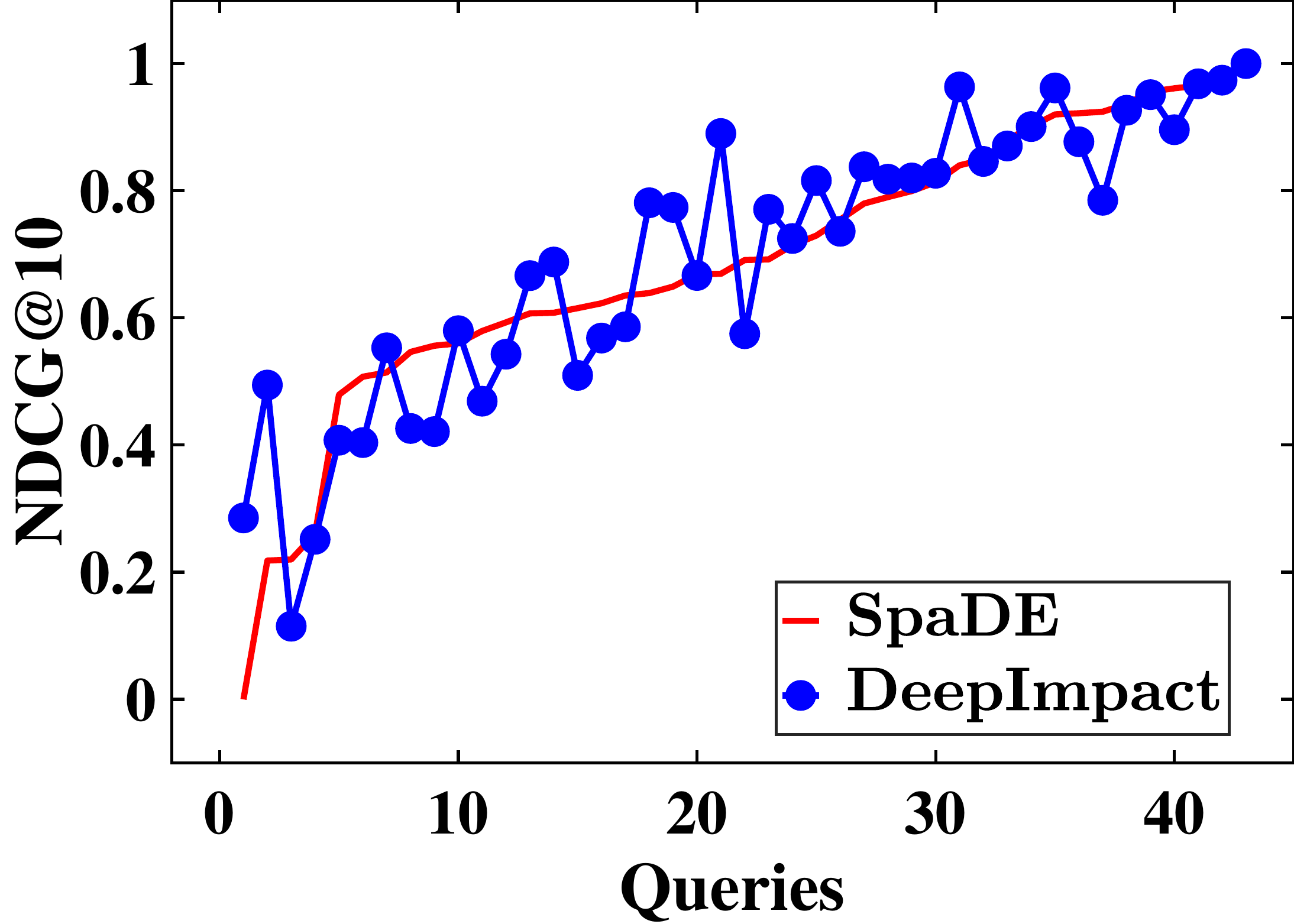} &
\includegraphics[width=0.22\textwidth]{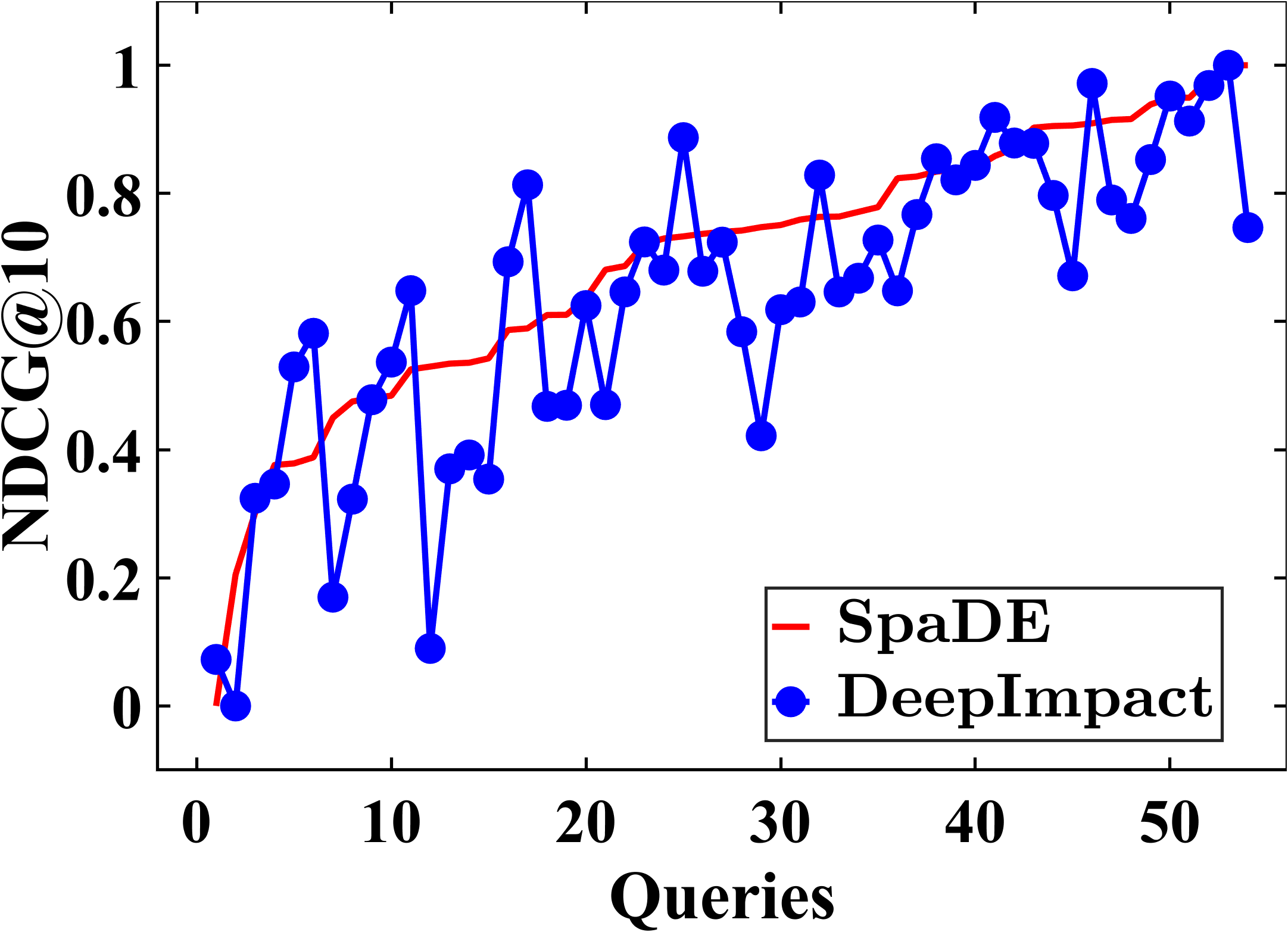} \\
(a) TREC DL 2019 & (b) TREC DL 2020 \\
\end{tabular}
\caption{NDCG@10 per query of SpaDE (\emph{k}=10) and DeepImpact~\cite{MalliaKTS21DeepImpact} on TREC DL 2019 and TREC DL 2020. We sort the queries by increasing order of NDCG@10 scores in SpaDE. If the blue points are above than the red line, it means that DeepImpact~\cite{MalliaKTS21DeepImpact} shows higher NDCG@10 score than SpaDE. For 51.2\% (22 out of 43) and 66.7\% (36 out of 54) queries on both datasets, SpaDE outperforms DeepImpact~\cite{MalliaKTS21DeepImpact}.}\label{fig:per_query_evaluation}
\vskip -0.1in
\end{figure}

\section{Detailed TREC DL Evaluation}\label{sec:app_pertopic_evaluation}

Figure~\ref{fig:per_query_evaluation} shows the NDCG@10 on each query of SpaDE and DeepImpact~\cite{MalliaKTS21DeepImpact}. While DeepImpact~\cite{MalliaKTS21DeepImpact} is better than SpaDE in TREC DL 2019, SpaDE is better than DeepImpact~\cite{MalliaKTS21DeepImpact} in TREC DL 2020. Our analysis found that incorrect queries of SpaDE on TREC DL 2019 mostly include relatively rare words. We conjecture that the term weighting component is difficult to assign high weighting scores on rare words because our learning scheme heavily depends on the document corpus.

\begin{table*}[t!]
\definecolor{pptred}{RGB}{192,0,0}
\definecolor{newyellow}{HTML}{FFE347}
\caption{Visualization of term weighting and document expansion by SpaDE (\emph{k}=10) and DeepImpact~\cite{MalliaKTS21DeepImpact} on TREC DL 2019 and 2020. The number in parentheses denotes the rank of the document. The tokenization process is simplified and only top-50 expanded words are visualized. \colorbox{newyellow!35}{Yellow shades} reflect the normalized term weights and query terms are boxed in \fcolorbox{pptred}{white}{\textbf{\textit{red}}}.}
\label{tab:heatmap}
\begin{tabular}{l}

% table 1
\begin{tabular}{>{\centering\arraybackslash} m{2cm} | m{7.2cm} m{7.2cm}}
\toprule
Query             & What is a nonconformity? earth science        \\ \midrule
Model             & DeepImpact~\cite{MalliaKTS21DeepImpact} (\#77) & SpaDE (\emph{k}=10) (\#27) \\ \midrule
Original document (Rel=3) & \includegraphics[width=7.2cm]{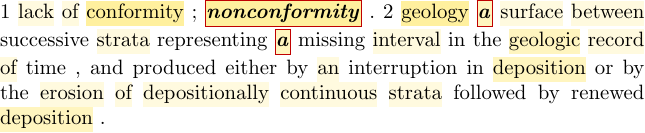} &   \includegraphics[width=7cm]{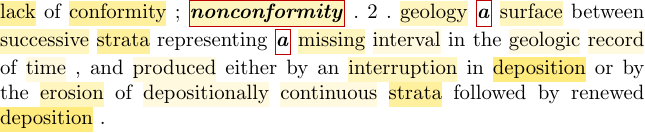}    \\ \midrule
Expanded terms & \includegraphics[width=7.2cm]{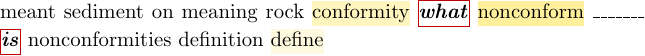} &    \includegraphics[width=7cm]{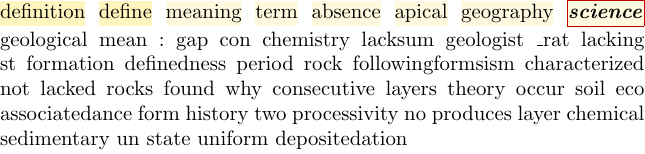}   \\ \bottomrule 
\end{tabular} \\ \\

% table 2
\begin{tabular}{>{\centering\arraybackslash} m{2cm} | m{7.2cm} m{7.2cm}}
\toprule
Query             &  dog day afternoon meaning       \\ \midrule
Model             & DeepImpact~\cite{MalliaKTS21DeepImpact} (\#26) & SpaDE (\emph{k}=10) (\#9) \\ \midrule
Original document (Rel=3) & \includegraphics[width=7.2cm]{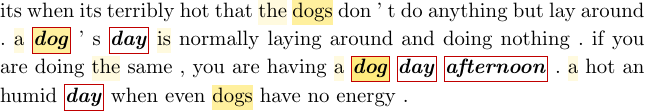} &   \includegraphics[width=7cm]{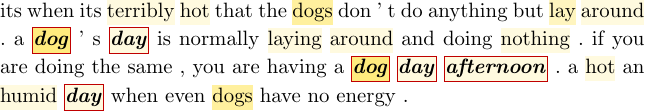}    \\ \midrule
Expanded terms & \includegraphics[width=7.2cm]{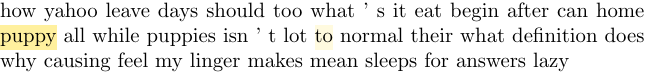} &    \includegraphics[width=7cm]{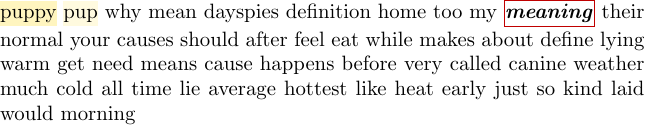}   \\ \bottomrule
\end{tabular} \\ \\

% table 3
\begin{tabular}{>{\centering\arraybackslash} m{2cm} | m{7.2cm} m{7.2cm}}
\toprule
Query             & what is the most popular food in switzerland         \\ \midrule
Model             & DeepImpact~\cite{MalliaKTS21DeepImpact} (\#9) & SpaDE (\emph{k}=10) (\#10) \\ \midrule
Original document (Rel=3) & \includegraphics[width=7.2cm]{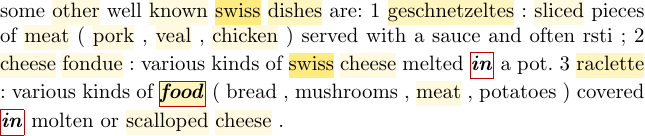} &   \includegraphics[width=7cm]{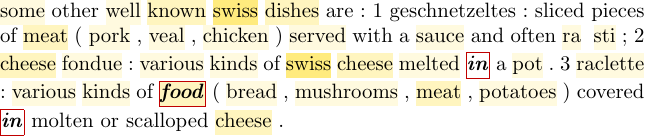}    \\ \midrule
Expanded terms & \includegraphics[width=7.2cm]{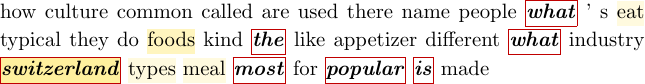} &    \includegraphics[width=7cm]{Figures/tab_heatmap_deepimpact_expand_3.pdf}   \\ \bottomrule
\end{tabular} \\

\end{tabular}
\end{table*}

\section{Visualization}\label{sec:app_visualization}

Table~\ref{tab:heatmap} shows the weighted document terms and expanded terms. SpaDE can identify important terms and expand new terms of SpaDE and DeepImpact~\cite{MalliaKTS21DeepImpact}. For the query \textit{"What is a nonconformity? earth science"}, SpaDE gives high scores for the important terms, \eg, \textit{"nonconformity"} and expanding the terms, \eg, \textit{"science"}, which are matched to the query. Owing to enriched document representations, SpaDE can rank a given relevant document in the 27th place, where the core terms of the query, \eg, \textit{"earth"}, \textit{"science"}, do not appear in the document. Meanwhile, even though DeepImpact~\cite{MalliaKTS21DeepImpact} identifies important document terms, the given document is ranked 77th since query terms are not properly expanded by docT5query~\cite{NogueiraL19docT5query}. SpaDE accurately expands the query terms by inferring the relevant words. For the query \textit{"Dog day afternoon meaning"}, SpaDE weights for the important terms, \eg, \textit{"dog", "day", "afternoon"}. Moreover, SpaDE successfully expands terms such as \textit{"warm", "weather", "hottest"}, which are highly related to the query. For the query \textit{"What is the most popular food in Switzerland"}, both DeepImpact and Spade give high weights for the important terms and expand relevant terms which are matched to the query, \eg, \textit{"Switzerland", "popular", "most"}.

\clearpage

\end{document}